\PassOptionsToPackage{unicode=true}{hyperref} 
\PassOptionsToPackage{hyphens}{url}
\documentclass[11pt,]{article}
\usepackage{lmodern}
\usepackage{setspace}
\usepackage{amssymb,amsmath}
\usepackage{ifxetex,ifluatex}
\usepackage{fixltx2e} 
\ifnum 0\ifxetex 1\fi\ifluatex 1\fi=0 
  \usepackage[T1]{fontenc}
  \usepackage[utf8]{inputenc}
  \usepackage{textcomp} 
\else 
  \usepackage{unicode-math}
  \defaultfontfeatures{Ligatures=TeX,Scale=MatchLowercase}
\fi
\IfFileExists{upquote.sty}{\usepackage{upquote}}{}
\IfFileExists{microtype.sty}{%
\usepackage[]{microtype}
\UseMicrotypeSet[protrusion]{basicmath} 
}{}
\IfFileExists{parskip.sty}{%
\usepackage{parskip}
}{
\setlength{\parindent}{0pt}
\setlength{\parskip}{6pt plus 2pt minus 1pt}
}
\usepackage{hyperref}
\hypersetup{
            pdfborder={0 0 0},
            breaklinks=true}
\usepackage[margin=1in]{geometry}
\usepackage{longtable,booktabs}
\IfFileExists{footnote.sty}{\usepackage{footnote}\makesavenoteenv{longtable}}{}
\usepackage{graphicx,grffile}
\makeatletter
\def\maxwidth{\ifdim\Gin@nat@width>\linewidth\linewidth\else\Gin@nat@width\fi}
\def\maxheight{\ifdim\Gin@nat@height>\textheight\textheight\else\Gin@nat@height\fi}
\makeatother
\setkeys{Gin}{width=\maxwidth,height=\maxheight,keepaspectratio}
\ifx\paragraph\undefined\else
\let\oldparagraph\paragraph
\renewcommand{\paragraph}[1]{\oldparagraph{#1}\mbox{}}
\fi
\ifx\subparagraph\undefined\else
\let\oldsubparagraph\subparagraph
\renewcommand{\subparagraph}[1]{\oldsubparagraph{#1}\mbox{}}
\fi

\makeatletter
\def\fps@figure{htbp}
\makeatother

\usepackage{rotating}

\usepackage{sidecap}
\makeatletter
\@ifpackageloaded{subfig}{}{\usepackage{subfig}}
\@ifpackageloaded{caption}{}{\usepackage{caption}}
\AtBeginDocument{%

}
\AtBeginDocument{%

}
\@ifpackageloaded{float}{}{\usepackage{float}}
\floatstyle{ruled}
\@ifundefined{c@chapter}{\newfloat{codelisting}{h}{lop}}{\newfloat{codelisting}{h}{lop}[chapter]}
\floatname{codelisting}{Listing}

\makeatother

\title{Separating the signal from the noise:\\
Evidence for deceleration in old-age death rates}
\author{Dennis M. Feehan\\
feehan@berkeley.edu}
\providecommand{\institute}[1]{}
\institute{UC Berkeley}
\date{March 27, 2018}

\begin{document}
\maketitle
\begin{abstract}
Widespread population aging has made it critical to understand death
rates at old ages. However, studying mortality at old ages is
challenging because the data are sparse: numbers of survivors and deaths
get smaller and smaller with age. We show how to address this challenge
by using principled model selection techniques to empirically evaluate
theoretical mortality models. We test nine different theoretical models
of old-age death rates by fitting them to 360 high-quality datasets on
cohort mortality above age 80. Models that allow for the possibility of
decelerating death rates tend to fit better than models that assume
exponentially increasing death rates. No single model is capable of
universally explaining observed old-age mortality patterns, but the
Log-Quadratic model most consistently predicts well. Patterns of model
fit differ by country and sex; we discuss possible mechanisms, including
sample size, period effects, and regional or cultural factors that may
be important keys to understanding patterns of old-age mortality. We
introduce a freely available R package that enables researchers to
extend our analysis to other models, age ranges, and data sources.
\end{abstract}

\newpage

\hypertarget{introduction}{%
\section{Introduction}\label{introduction}}

In the developed world, life expectancy at birth continues to rise and
fertility levels are low. At the same time, much of the developing world
has seen declines in birthrates and improvements in child and adult
survival (United Nations Population Division 2015). Together, these
trends have led demographers to forecast widespread and rapid population
aging (Gerland et al. 2014). This shifting demographic landscape raises
critical scientific and policy questions about the nature of human
longevity (Kinsella et al. 2005; Martin and Preston 1994). To understand
this phenomenon, researchers have proposed several different theories
that aim to explain and predict patterns of death rates at advanced
ages.\footnote{In this paper, we use the term `advanced ages' or `oldest
  ages' to refer to ages over 80.} This study carefully assesses the
amount of empirical support that these proposed theories have.

Understanding theoretical models of old-age mortality is important for
several reasons. First, these models are critical to progress in
answering a key question in the science of ageing: can we expect
lifespans to increase indefinitely, or is there some upper limit at
which our biology renders continued improvements essentially impossible
(Dong et al. 2016; Lenart and Vaupel 2017)? To answer this question,
researchers need evidence about empirical regularities in old-age death
rates. Unfortunately, the sparse numbers of deaths at advanced ages has
made establishing these empirical regularities quite difficult. For
example, although it is widely accepted that death rates accelerate over
the course of middle and early-old ages, there is less agreement about
what happens to mortality at advanced ages: while some models of old-age
mortality imply that death rates will continue to increase at oldest
ages (Dong et al. 2016; Gompertz 1825; Makeham 1860), other models
predict that death rates will decelerate or even plateau (L. A. Gavrilov
and Gavrilova 2011; Horiuchi and Wilmoth 1998; A. Thatcher et al. 1998).
Second, many policymakers and planners need accurate, mathematical
summaries of death rates at advanced ages in order to produce forecasts
and projections. For example, old-age mortality forecasts are a critical
input to planning for health care and pension systems (Bongaarts 2005;
Tabeau et al. 2001). Finally, understanding old-age survival patterns is
important for a number of other research questions of great relevance to
sociology, economics, and public policy. For example, a mathematical
description of old-age mortality is needed to build a behavioral model
of how increases in longevity are related to changes in savings and
investment behavior (Sheshinski 2007).

The remainder of this article begins by briefly reviewing the theory
behind nine leading models of old-age mortality
(Section~\ref{sec:background}). We then introduce the dataset, which
consists of high-quality information on population and deaths by age for
360 cohorts selected from the Kannisto-Thatcher database on old-age
mortality (Section~\ref{sec:data}). Next, we discuss how to measure the
relative amount of empirical support the various theories have
(Section~\ref{sec:methods}). This topic is critical because deaths at
old ages are very rare, so careful attention must be paid to how models
are fit and assessed. We then turn to a substantive analysis of the
resulting model fits (Section~\ref{sec:results}). We find that models
that allow for the possibility of decelerating death rates tend to fit
better than models that assume continual exponential increase. The
Log-Quadratic model performs best, but no single model can be said to
adequately capture patterns of old-age mortality across all of the
cohorts in our sample; instead, patterns of model fit vary considerably
by country. We relate our findings to previous research and discuss
several mechanisms that may be important determinants of old-age
mortality patterns, including period effects, cultural factors, and
country-specific differences in sample size. The paper concludes with a
discussion and an outline of topics for future research
(Section~\ref{sec:conclusion}). As a companion to the article, we
introduce a freely available R package that can be used to extend our
analysis to other age ranges, models, and datasets.

\hypertarget{sec:background}{%
\section{Background and theory}\label{sec:background}}

\hypertarget{mortality-models}{%
\subsection{Mortality models}\label{mortality-models}}

Theoretical models of old-age mortality can be expressed in terms of (i)
an individual hazard function and (ii) an aggregation method. Together,
these two components lead to a \emph{population hazard function}
(Figure~\ref{fig:theory-and-agg}; Appendix \ref{sec:ap-heterogeneity}).
For a group of people, the \emph{population hazard function} is defined
as:

\begin{equation}
    \mu(z) = -\frac{d \log S(z)}{d z} = - \frac{1}{S(z)} \frac{d S(z)}{d z}. 
\label{eq:popnhaz}\end{equation}

\noindent where \(S(z)\) is the \emph{population survival function},
\emph{i.e.} the proportion of group members that survives to exact age
\(z\).

Given a model for the population hazard, there is a mathematical
relationship that links the hazard function \(\mu(z)\) and the
probability of dying between ages \(z\) and \(z+k\), conditional on
surviving to age \(z\):

\begin{equation}
\pi(z,z+k) = 1 - \exp\left( - \int_{z}^{z+k} \mu(x) dx\right).
\label{eq:deathprob}\end{equation}

\noindent Equation~\ref{eq:deathprob} shows that a population hazard
function can be converted into expected numbers of cohort deaths by age.
A theory that has been expressed as a hazard function can thus be tested
by comparing predicted numbers of deaths by age to empirically observed
numbers of deaths by age.

Scientifically, it is important to understand both (1) how well a
particular theory fits empirical data in an absolute sense; and (2) how
well a theory fits empirical data \emph{relative to other theories}.
Understanding how well a particular theory fits the data in an absolute
sense is critical to its plausibility: a theory that is entirely
inconsistent with empirical evidence must be discarded. Understanding
how well a theory fits relative to other theories is critical to
progress: even a theory that produces adequate fits to empirical data
may be replaced by an alternate theory that consistently fits the data
better. Relative comparisons of theories might also reveal that there is
no consistent pattern to which theory fits data best, suggesting that
multiple theories may have merit; that different theoretical mechanisms
are important for different subgroups; or that adequate explanatory
theories have not yet been proposed.

\hypertarget{fig:theory-and-agg}{%
\begin{figure}
\centering
\includegraphics{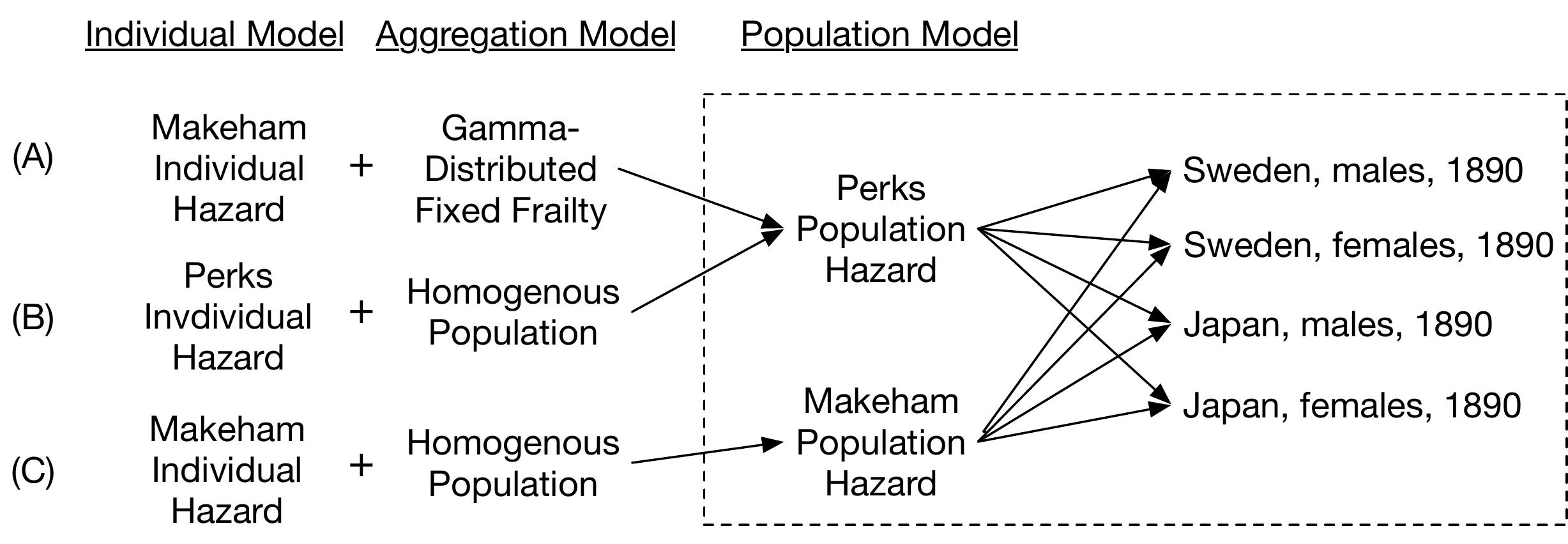}
\caption{Illustrative example of three different possible mortality
theories and their empirical predictions. The left-hand column shows
different hazards that individuals might face. The next column shows
examples of different ways that these individual hazards might aggregate
up into the population hazard. Two different theoretical models, (A) and
(B) lead to the same empirical prediction about cohort deaths by age (A.
Yashin et al. 1994). Thus, models (A) and (B) cannot be distinguished
without additional information. A third model, (C), makes a different
empirical prediction. Thus, information on cohort deaths by age can be
used to determine whether model (C) is better or worse than models (A)
and (B). The dashed line shows what can be understood from analyzing
cohort deaths by age, which is the focus of the analysis
below.}\label{fig:theory-and-agg}
\end{figure}
}

\hypertarget{functional-forms-for-the-population-hazard}{%
\subsection{Functional Forms for the Population
Hazard}\label{functional-forms-for-the-population-hazard}}

The most prominent models of old-age mortality have been expressed as
hazard functions (Table~\ref{tbl:hazfns}). We now briefly review the
theory associated with each of these models\footnote{Note that, although
  our focus is on old-age mortality, these models can also be applied to
  other age ranges.}. These models can be described in four sets.

The first set contains models that can be considered generalizations of
the Gompertz model. The Gompertz model is the oldest mathematical
description of mortality; it was first proposed as a descriptive theory
that could closely capture empirically observed risks of death, which
increase rapidly across adult ages (Gompertz 1825). Researchers
following Gompertz developed mechanism-based probability models that can
produce Gompertzian mortality dynamics (Strehler and Mildvan 1960).
Gompertz is considered to provide a good, stylized description of adult
mortality throughout the middle to early-old age range, but some
researchers argue that it fails to capture the dynamics of oldest-age
mortality patterns (e.g., Vaupel et al. 1998). The Makeham model was
designed to generalize the Gompertz model by adding a parameter that
captures age-invariant `background' mortality; this background mortality
could, for example, describe everyone's risk of death due to
environmental factors that do not vary with age (Makeham 1860). The
Log-Quadratic model has emerged from biodemographic investigations of
lifespan and aging rates, and can be seen as a generalization of the
Makeham model that adds a parameter which allows increases in death
rates to decelerate at advanced ages (Coale and Kisker 1990; Steinsaltz
and Wachter 2006; J. R. Wilmoth 1995).

The second set contains the Weibull model, which can be motivated by
drawing analogies between human biology and theories of reliability
rooted in engineering and manufacturing. These reliability theories
describe systems composed of multiple components. Each component has
some time-dependent risk of failure, independent of the other components
(Weibull 1951). The Weibull model would make sense if, for example, the
human body could be accurately modeled as a set of independent organs,
with death resulting when any of the organs fails.

The third set contains the Logistic, Beard, Perks, and Kannisto models.
These four models arose from research on heterogeneity in mortality;
they result from postulating a distribution over individual hazard
functions, leading to a population hazard that is a continuous mixture
of these individual hazards (R. E Beard 1959; Perks 1932; Vaupel et al.
1979). For example, the Beard model can be derived by assuming that
individuals face a Gompertz hazard and that each individual's hazard has
been scaled by a Gamma-distributed variate often called `frailty' (R. E
Beard 1959; Horiuchi and Wilmoth 1998; K. G. Manton et al. 1981). The
Kannisto model is unique in this set because it was not derived from
first principles; instead, it was proposed as a heuristic two-parameter
approximation of the more complex heterogeneity models (Himes et al.
1994; A. Thatcher et al. 1998).

Finally, the fourth set contains the Lynch-Brown model, which was
devised to describe and interpret observed crossovers in death rates
between Black and White populations in the United States (S. M. Lynch
and Brown 2001; Scott M. Lynch et al. 2003).

\hypertarget{tbl:hazfns}{}
\begin{longtable}[]{@{}llll@{}}
\caption{\label{tbl:hazfns}The functional forms for the hazard of death
at advanced ages considered in this analysis. In the functions listed
above, \(z\) is age and \(\mu(z)\) is the population hazard function. L.
A Gavrilov and Gavrilova (1991) and A. Thatcher et al. (1998) contain
helpful discussions about many of these functional
forms.}\tabularnewline
\toprule
\begin{minipage}[b]{0.18\columnwidth}\raggedright
Name\strut
\end{minipage} & \begin{minipage}[b]{0.14\columnwidth}\raggedright
Parameters\strut
\end{minipage} & \begin{minipage}[b]{0.38\columnwidth}\raggedright
Function\strut
\end{minipage} & \begin{minipage}[b]{0.19\columnwidth}\raggedright
Sample references\strut
\end{minipage}\tabularnewline
\midrule
\endfirsthead
\toprule
\begin{minipage}[b]{0.18\columnwidth}\raggedright
Name\strut
\end{minipage} & \begin{minipage}[b]{0.14\columnwidth}\raggedright
Parameters\strut
\end{minipage} & \begin{minipage}[b]{0.38\columnwidth}\raggedright
Function\strut
\end{minipage} & \begin{minipage}[b]{0.19\columnwidth}\raggedright
Sample references\strut
\end{minipage}\tabularnewline
\midrule
\endhead
\begin{minipage}[t]{0.18\columnwidth}\raggedright
Gompertz\strut
\end{minipage} & \begin{minipage}[t]{0.14\columnwidth}\raggedright
\(\alpha\), \(\beta\)\strut
\end{minipage} & \begin{minipage}[t]{0.38\columnwidth}\raggedright
\(\mu(z) = \alpha \exp(\beta z)\)\strut
\end{minipage} & \begin{minipage}[t]{0.19\columnwidth}\raggedright
Gompertz (1825)\strut
\end{minipage}\tabularnewline
\begin{minipage}[t]{0.18\columnwidth}\raggedright
Kannisto\strut
\end{minipage} & \begin{minipage}[t]{0.14\columnwidth}\raggedright
\(\alpha\), \(\beta\)\strut
\end{minipage} & \begin{minipage}[t]{0.38\columnwidth}\raggedright
\(\mu(z) = \frac{\alpha \exp(\beta z) } {1 + \exp(\beta z)}\)\strut
\end{minipage} & \begin{minipage}[t]{0.19\columnwidth}\raggedright
A. Thatcher et al. (1998)\strut
\end{minipage}\tabularnewline
\begin{minipage}[t]{0.18\columnwidth}\raggedright
Weibull\strut
\end{minipage} & \begin{minipage}[t]{0.14\columnwidth}\raggedright
\(\alpha\), \(\beta\)\strut
\end{minipage} & \begin{minipage}[t]{0.38\columnwidth}\raggedright
\(\mu(z) = \alpha z^{\beta-1}\)\strut
\end{minipage} & \begin{minipage}[t]{0.19\columnwidth}\raggedright
Weibull (1951)\strut
\end{minipage}\tabularnewline
\begin{minipage}[t]{0.18\columnwidth}\raggedright
Makeham\strut
\end{minipage} & \begin{minipage}[t]{0.14\columnwidth}\raggedright
\(\alpha\), \(\beta\), \(\gamma\)\strut
\end{minipage} & \begin{minipage}[t]{0.38\columnwidth}\raggedright
\(\mu(z) = \gamma + \alpha \exp(\beta z)\)\strut
\end{minipage} & \begin{minipage}[t]{0.19\columnwidth}\raggedright
Makeham (1860)\strut
\end{minipage}\tabularnewline
\begin{minipage}[t]{0.18\columnwidth}\raggedright
Beard / Gamma-Gompertz\strut
\end{minipage} & \begin{minipage}[t]{0.14\columnwidth}\raggedright
\(\alpha\), \(\beta\), \(\delta\),\strut
\end{minipage} & \begin{minipage}[t]{0.38\columnwidth}\raggedright
\(\mu(z) = \frac{\alpha \exp(\beta z)} {1 + \delta \exp(\beta z)}\)\strut
\end{minipage} & \begin{minipage}[t]{0.19\columnwidth}\raggedright
R. E Beard (1959), Horiuchi and Wilmoth (1998)\strut
\end{minipage}\tabularnewline
\begin{minipage}[t]{0.18\columnwidth}\raggedright
Log-Quadratic\strut
\end{minipage} & \begin{minipage}[t]{0.14\columnwidth}\raggedright
\(\alpha\), \(\beta\), \(\gamma\)\strut
\end{minipage} & \begin{minipage}[t]{0.38\columnwidth}\raggedright
\(\mu(z) = \exp(\alpha +\beta z + \gamma z^2)\)\strut
\end{minipage} & \begin{minipage}[t]{0.19\columnwidth}\raggedright
Coale and Kisker (1990), J. R. Wilmoth (1995) Steinsaltz and Wachter
(2006)\strut
\end{minipage}\tabularnewline
\begin{minipage}[t]{0.18\columnwidth}\raggedright
Logistic\strut
\end{minipage} & \begin{minipage}[t]{0.14\columnwidth}\raggedright
\(\alpha\), \(\beta\), \(\gamma\), \(\delta\)\strut
\end{minipage} & \begin{minipage}[t]{0.38\columnwidth}\raggedright
\(\mu(z) = \gamma + \frac{\alpha \exp(\beta z)} {1 + \delta \exp(\beta z)}\)\strut
\end{minipage} & \begin{minipage}[t]{0.19\columnwidth}\raggedright
Perks (1932), Robert E. Beard (1971)\strut
\end{minipage}\tabularnewline
\begin{minipage}[t]{0.18\columnwidth}\raggedright
Perks / Gamma-Makeham\strut
\end{minipage} & \begin{minipage}[t]{0.14\columnwidth}\raggedright
\(\alpha\), \(\beta\), \(\gamma\), \(\delta\)\strut
\end{minipage} & \begin{minipage}[t]{0.38\columnwidth}\raggedright
\(\mu(z) = \frac{\gamma + \alpha \exp(\beta z)} {1 + \delta \exp(\beta z)}\)\strut
\end{minipage} & \begin{minipage}[t]{0.19\columnwidth}\raggedright
Perks (1932), R. E Beard (1959), Horiuchi and Wilmoth (1998)\strut
\end{minipage}\tabularnewline
\begin{minipage}[t]{0.18\columnwidth}\raggedright
Lynch-Brown\strut
\end{minipage} & \begin{minipage}[t]{0.14\columnwidth}\raggedright
\(\alpha\), \(\beta\), \(\gamma\), \(\delta\)\strut
\end{minipage} & \begin{minipage}[t]{0.38\columnwidth}\raggedright
\(\mu(z) = \alpha + \beta \text{arctan} \{\gamma(z-\delta)\}\)\strut
\end{minipage} & \begin{minipage}[t]{0.19\columnwidth}\raggedright
S. M. Lynch and Brown (2001)\strut
\end{minipage}\tabularnewline
\bottomrule
\end{longtable}

To recap: many theories have been proposed to explain the patterns of
mortality at oldest ages. Each theory corresponds to a population hazard
function which mathematically describes the predicted risk of death by
age for a group of people. Population hazards can be converted into
expected numbers of deaths by age, and predicted deaths by age can then
be compared to observed deaths by age to determine how well each theory
agrees with empirical evidence. Appendix \ref{sec:hazards} provides
additional details and references to the literature.

\hypertarget{sec:data}{%
\section{Data}\label{sec:data}}

In order to empirically test the models in Table~\ref{tbl:hazfns}, we
fit each model to data from the Kannisto-Thatcher (K-T) database on Old
Age Mortality. The K-T database is a carefully curated collection of
data on mortality at advanced ages, with information on deaths and
exposure above age 80 for 35 countries, with some of the Scandinavian
data going back to the mid-18th century (Kannisto et al. 1994; MPIDR
2014). Since data quality for mortality at advanced ages is a serious
concern (e.g., Black et al. 2017), we only analyze the subset of
country-cohorts that were found to be of high quality by the expert
review of Jdanov et al. (2008)\footnote{Jdanov et al. (2008) developed
  methods to quantify several different ways that age patterns of deaths
  and population counts can be irregular. The researchers then
  systematically applied their methods to each cohort in the K-T
  database, producing comparable indicators for data quality across
  cohorts. Finally, Jdanov et al. (2008) summarized their results using
  hierarchical clustering, producing four tiers of data quality: best,
  acceptable, conditionally acceptable, and weak.}. Specifically, we
retained the cohort data from all countries where more than half of the
years were of the highest quality, and the remaining years were of the
second-highest quality in the assessment the authors provided.
Furthermore, we only selected countries and time periods where deaths
were reported by single year of age up to 104. This leaves data from
Denmark, France, West Germany, Italy, Japan, the Netherlands, Sweden,
and Switzerland. In total, the analysis dataset has 360 cohorts of data
on survivors and deaths by sex and single year of age from 10 different
countries. Table~\ref{tbl:kt-data} describes the years and cohorts in
the analysis dataset, and Appendix \ref{sec:ap-data} has more detail on
how the analysis sample was chosen.

\hypertarget{tbl:kt-data}{}
\begin{longtable}[]{@{}cccccc@{}}
\caption{\label{tbl:kt-data}Cohorts from the Kannisto-Thatcher Database
on Old Age Mortality used in this analysis. Only cohorts with the
highest-quality data and deaths reported at least up to age 105 by
single year of age are included. The final dataset has data for 360
country-sex-cohorts from 10 countries. Jdanov et al. (2008) has a
detailed discussion of data quality and Appendix \ref{sec:ap-data} has
more information about how the analysis dataset was
constructed.}\tabularnewline
\toprule
\begin{minipage}[b]{0.13\columnwidth}\centering
Country\strut
\end{minipage} & \begin{minipage}[b]{0.13\columnwidth}\centering
Cohort start\strut
\end{minipage} & \begin{minipage}[b]{0.11\columnwidth}\centering
Cohort end\strut
\end{minipage} & \begin{minipage}[b]{0.17\columnwidth}\centering
Number of cohorts\strut
\end{minipage} & \begin{minipage}[b]{0.15\columnwidth}\centering
Avg. size (80)\strut
\end{minipage} & \begin{minipage}[b]{0.15\columnwidth}\centering
Avg. size (95)\strut
\end{minipage}\tabularnewline
\midrule
\endfirsthead
\toprule
\begin{minipage}[b]{0.13\columnwidth}\centering
Country\strut
\end{minipage} & \begin{minipage}[b]{0.13\columnwidth}\centering
Cohort start\strut
\end{minipage} & \begin{minipage}[b]{0.11\columnwidth}\centering
Cohort end\strut
\end{minipage} & \begin{minipage}[b]{0.17\columnwidth}\centering
Number of cohorts\strut
\end{minipage} & \begin{minipage}[b]{0.15\columnwidth}\centering
Avg. size (80)\strut
\end{minipage} & \begin{minipage}[b]{0.15\columnwidth}\centering
Avg. size (95)\strut
\end{minipage}\tabularnewline
\midrule
\endhead
\begin{minipage}[t]{0.13\columnwidth}\centering
Sweden\strut
\end{minipage} & \begin{minipage}[t]{0.13\columnwidth}\centering
1821\strut
\end{minipage} & \begin{minipage}[t]{0.11\columnwidth}\centering
1895\strut
\end{minipage} & \begin{minipage}[t]{0.17\columnwidth}\centering
150\strut
\end{minipage} & \begin{minipage}[t]{0.15\columnwidth}\centering
19,567\strut
\end{minipage} & \begin{minipage}[t]{0.15\columnwidth}\centering
1,088\strut
\end{minipage}\tabularnewline
\begin{minipage}[t]{0.13\columnwidth}\centering
France\strut
\end{minipage} & \begin{minipage}[t]{0.13\columnwidth}\centering
1866\strut
\end{minipage} & \begin{minipage}[t]{0.11\columnwidth}\centering
1892\strut
\end{minipage} & \begin{minipage}[t]{0.17\columnwidth}\centering
54\strut
\end{minipage} & \begin{minipage}[t]{0.15\columnwidth}\centering
159,094\strut
\end{minipage} & \begin{minipage}[t]{0.15\columnwidth}\centering
10,678\strut
\end{minipage}\tabularnewline
\begin{minipage}[t]{0.13\columnwidth}\centering
Netherlands\strut
\end{minipage} & \begin{minipage}[t]{0.13\columnwidth}\centering
1871\strut
\end{minipage} & \begin{minipage}[t]{0.11\columnwidth}\centering
1895\strut
\end{minipage} & \begin{minipage}[t]{0.17\columnwidth}\centering
50\strut
\end{minipage} & \begin{minipage}[t]{0.15\columnwidth}\centering
32,203\strut
\end{minipage} & \begin{minipage}[t]{0.15\columnwidth}\centering
2,640\strut
\end{minipage}\tabularnewline
\begin{minipage}[t]{0.13\columnwidth}\centering
Denmark\strut
\end{minipage} & \begin{minipage}[t]{0.13\columnwidth}\centering
1881\strut
\end{minipage} & \begin{minipage}[t]{0.11\columnwidth}\centering
1895\strut
\end{minipage} & \begin{minipage}[t]{0.17\columnwidth}\centering
30\strut
\end{minipage} & \begin{minipage}[t]{0.15\columnwidth}\centering
16,836\strut
\end{minipage} & \begin{minipage}[t]{0.15\columnwidth}\centering
1,478\strut
\end{minipage}\tabularnewline
\begin{minipage}[t]{0.13\columnwidth}\centering
Italy\strut
\end{minipage} & \begin{minipage}[t]{0.13\columnwidth}\centering
1881\strut
\end{minipage} & \begin{minipage}[t]{0.11\columnwidth}\centering
1895\strut
\end{minipage} & \begin{minipage}[t]{0.17\columnwidth}\centering
30\strut
\end{minipage} & \begin{minipage}[t]{0.15\columnwidth}\centering
159,915\strut
\end{minipage} & \begin{minipage}[t]{0.15\columnwidth}\centering
10,872\strut
\end{minipage}\tabularnewline
\begin{minipage}[t]{0.13\columnwidth}\centering
West Germany\strut
\end{minipage} & \begin{minipage}[t]{0.13\columnwidth}\centering
1891\strut
\end{minipage} & \begin{minipage}[t]{0.11\columnwidth}\centering
1895\strut
\end{minipage} & \begin{minipage}[t]{0.17\columnwidth}\centering
10\strut
\end{minipage} & \begin{minipage}[t]{0.15\columnwidth}\centering
227,391\strut
\end{minipage} & \begin{minipage}[t]{0.15\columnwidth}\centering
16,170\strut
\end{minipage}\tabularnewline
\begin{minipage}[t]{0.13\columnwidth}\centering
Japan\strut
\end{minipage} & \begin{minipage}[t]{0.13\columnwidth}\centering
1891\strut
\end{minipage} & \begin{minipage}[t]{0.11\columnwidth}\centering
1895\strut
\end{minipage} & \begin{minipage}[t]{0.17\columnwidth}\centering
10\strut
\end{minipage} & \begin{minipage}[t]{0.15\columnwidth}\centering
204,541\strut
\end{minipage} & \begin{minipage}[t]{0.15\columnwidth}\centering
18,361\strut
\end{minipage}\tabularnewline
\begin{minipage}[t]{0.13\columnwidth}\centering
Scotland\strut
\end{minipage} & \begin{minipage}[t]{0.13\columnwidth}\centering
1891\strut
\end{minipage} & \begin{minipage}[t]{0.11\columnwidth}\centering
1895\strut
\end{minipage} & \begin{minipage}[t]{0.17\columnwidth}\centering
10\strut
\end{minipage} & \begin{minipage}[t]{0.15\columnwidth}\centering
18,258\strut
\end{minipage} & \begin{minipage}[t]{0.15\columnwidth}\centering
1,468\strut
\end{minipage}\tabularnewline
\begin{minipage}[t]{0.13\columnwidth}\centering
Switzerland\strut
\end{minipage} & \begin{minipage}[t]{0.13\columnwidth}\centering
1891\strut
\end{minipage} & \begin{minipage}[t]{0.11\columnwidth}\centering
1895\strut
\end{minipage} & \begin{minipage}[t]{0.17\columnwidth}\centering
10\strut
\end{minipage} & \begin{minipage}[t]{0.15\columnwidth}\centering
21,349\strut
\end{minipage} & \begin{minipage}[t]{0.15\columnwidth}\centering
2,021\strut
\end{minipage}\tabularnewline
\begin{minipage}[t]{0.13\columnwidth}\centering
Belgium\strut
\end{minipage} & \begin{minipage}[t]{0.13\columnwidth}\centering
1893\strut
\end{minipage} & \begin{minipage}[t]{0.11\columnwidth}\centering
1895\strut
\end{minipage} & \begin{minipage}[t]{0.17\columnwidth}\centering
6\strut
\end{minipage} & \begin{minipage}[t]{0.15\columnwidth}\centering
38,416\strut
\end{minipage} & \begin{minipage}[t]{0.15\columnwidth}\centering
2,910\strut
\end{minipage}\tabularnewline
\bottomrule
\end{longtable}

\hypertarget{sec:methods}{%
\section{Methods}\label{sec:methods}}

\hypertarget{likelihood-and-estimation}{%
\subsection{Likelihood and estimation}\label{likelihood-and-estimation}}

If \(N_z\) people from a cohort survive to exact age \(z\), and all of
the members of the cohort face the same hazard \(\mu(z)\), then cohort
deaths between exact ages \(z\) and \(z+1\) are distributed binomially:

\[
D_z \sim \text{Binomial}(N_z, \pi(z, z+1)),
\]

\noindent where \(D_z\) is the number of deaths between ages \(z\) and
\(z+1\), and \(\pi(z,z+1)\) is the probability of dying between ages
\(z\) and \(z+1\) (Chiang 1960).

\noindent We fit each model to each cohort using the method of maximum
likelihood. The likelihood for an observed sequence of deaths
\(\mathbf{D} = D_1, D_2, \dots\) and survivors to each age,
\(\mathbf{N} = N_1, N_2, \dots\) is then

\begin{equation}
Pr(\mathbf{D}| \mathbf{z}, \theta,\mathbf{N}) = \prod_z {\binom{N_z}{D_z}} \pi(z,\theta)^{D_z} (1-\pi(z,\theta))^{(N_z - D_z)}.
\label{eq:likelihood}\end{equation}

Taking logs yields

\begin{equation}
  ll(\mathbf{D}| \mathbf{z}, \theta, \mathbf{N}) = K + \sum_z\left[
    D_z\log(\pi(z, \theta)) + (N_z-D_z)\log(1-\pi(z,
    \theta))\right],
\label{eq:log-likelihood}\end{equation}

\noindent where \(K\) is a constant that does not vary with \(\theta\).
In order to fit a particular hazard model to a dataset
(\(\mathbf{D}, \mathbf{N})\) from a particular cohort,
Equation~\ref{eq:log-likelihood} is maximized as a function of
\(\theta\). The computation required to perform this maximization is not
trivial; therefore, in Appendix \ref{sec:ap-model} and
\ref{sec:ap-estimation} we provide a more detailed description of the
methods we use to fit each cohort dataset and we introduce the freely
available R package \texttt{mortfit} which other researchers can use to
develop and fit their own mortality models\footnote{We also make use of
  several R packages, including Daróczi and Tsegelskyi (2015), R Core
  Team (2014), H. Wickham (2009), Slowikowski and Irisson (2016),
  Schloerke et al. (2014), and H. Wickham et al. (2016).}.

There are several advantages to fitting mortality hazards by building a
likelihood-based model for the observed deaths by age. First, it is
uncommon for people to survive to old ages, making absolute numbers of
deaths and survivors small and rapidly decreasing by age. Fitting
hazards using maximum likelihood naturally accounts for the amount of
data available at each age; intuitively, older ages with fewer survivors
and deaths have less information about likely parameter values, and thus
less influence on parameter estimates than younger ages where lots of
data are observed. A different approach, such as fitting a regression to
observed central death rates, can result in parameter estimates that are
heavily influenced by noisy observations at the oldest ages. Several
studies have argued for the benefits of likelihood-based estimates of
mortality models (K. G. Manton et al. 1981; Pletcher 1999; Promislow et
al. 1999; Steinsaltz 2005; Wang et al. 1998). Second, fitting mortality
hazards using likelihood-based inference enables the use of principled
model selection techniques to help determine which model is most
consistent with the data. The theory for one important technique, the
AIC, relies upon likelihood-based inference. (Below, we discuss why
these principled model selection techniques are important.) Finally, we
model the counts of deaths by age because those are the data that are
actually observed; modeling another quantity, such as central death
rates, would require making additional assumptions.

\hypertarget{model-selection}{%
\subsection{Model selection}\label{model-selection}}

Now we turn to methods that can be used to assess how well each fitted
model explains observed deaths in each cohort. One approach is to
directly compare the observed number of deaths in a cohort to the
model's predicted number of deaths. For example, we can compute the sum
of squared errors in the estimated number of deaths for each
model-country-sex-year:

\begin{equation}
  \text{SSE} = \sum_z \left( \hat{D_z} - D_z \right)^2,
\label{eq:sse}\end{equation}

where \(z\) ranges over the ages in the dataset. This quantity provides
a natural measurement of each model's \emph{accuracy}---\emph{i.e.}, how
close each model's predictions come to the observed numbers of deaths,
in absolute terms.

The statistical literature on model selection has revealed that accuracy
is only one of two important factors in assessing model fit. The second
important factor is \emph{generalizability}, which is the extent to
which a model's performance on the observed dataset can be expected to
be replicated on future datasets that arise from the same data
generating process. \emph{Overfitting} is a crucial consideration here
(Burnham and Anderson 2003; Claeskens and Hjort 2008). Models with many
free parameters are more flexible, meaning that they can bend to fit the
shape of observed patterns in the data; thus, models with many free
parameters can be expected to fit any particular dataset closely. This
flexibility could be advantageous or disadvantageous: a more flexible
model would be able to capture more complex relationships between age
and the risk of death; however, a more flexible model could also
overfit---i.e., a flexible model might pick up noisy artefacts due to
small sample sizes at old ages.

Model selection techniques address the problem of overfitting by
penalizing a model's \emph{complexity}--i.e., the number of free
parameters it has. Thus, model selection techniques trade off accuracy
and model complexity. Since the SSE has no penalty for model complexity,
models that overfit can nonetheless appear to perform very well
according to SSE. In order to account for both model accuracy and model
generalizability, we evaluate model fits by focusing on an alternative
metric called Akaike's Information Criterion (AIC). The AIC is
essentially a penalized log-likelihood, where the penalty is a function
of the number of parameters being estimated in the model:

\begin{equation}
  \text{AIC} = -2  \mathfrak{L} + 2k,
\label{eq:aic}\end{equation}

where \(\mathfrak{L}\) is the value of the maximized likelihood from
Equation~\ref{eq:log-likelihood}, and \(k\) is the number of parameters
being estimated. Although Equation~\ref{eq:aic} looks simple, it has
been motivated by extensive theory. This theory shows that the AIC
selects the candidate model that minimizes the expected Kullback-Leibler
distance between the distribution of data implied by the model and the
one that generated the data (Akaike 1974; Burnham and Anderson 2003;
Claeskens and Hjort 2008). The derivation of the AIC reveals that the
second term (\(2k\)) is a bias adjustment that corrects for overfitting;
thus, the AIC results are an indication of how well each hazard function
trades off the number of parameters estimated with the accuracy of its
fit to the data.

The derivation of the AIC reveals that absolute AIC values are not
interpretable; however, two models can be compared by examining the
difference between their AICs. Thus, the results below focus on a
quantity called \(\Delta\text{AIC}\). For a particular model fit,
\(\Delta\text{AIC}\) is the difference between (1) the AIC value
attained by the model and (2) the AIC value obtained for the model that
best fit the dataset. Thus, for the best-fitting model,
\(\Delta\text{AIC}\) is 0; for other models, \(\Delta\text{AIC}\) will
be greater than 0. The closer \(\Delta\text{AIC}\) is to 0, the better
its fit.

The results below focus on the AIC because it accounts for the tension
between model complexity and accuracy; because it is motivated by
rigorous statistical theory; and because it is widely used in the
natural and social sciences. Alternative theoretical perspectives have
been used to develop other principled model selection criteria. Appendix
\ref{sec:ap-results} describes two of these alternatives-- the Bayesian
Information Criterion and \(K\)-fold cross-validation-- and compares the
results based on the AIC to analogous results using these two
alternative methods. We believe that the theory that motivates the AIC
is most relevant to our goal in this study, but more detailed
discussions of model selection, including comparisons between the AIC
and BIC, can be found in Claeskens and Hjort (2008); Burnham and
Anderson (2003); Burnham and Anderson (2004); Hoeting et al. (1999);
Friedman et al. (2009); and Kass and Raftery (1995).

\hypertarget{sec:results}{%
\section{Results}\label{sec:results}}

Each model in Table~\ref{tbl:hazfns} was fit separately for males and
females in each cohort in Table~\ref{tbl:kt-data}. To build intuition,
we first illustrate the results for a single cohort. Next, we examine
the results for the entire dataset and for specific countries.

\hypertarget{an-illustrative-example}{%
\subsection{An illustrative example}\label{an-illustrative-example}}

Figure~\ref{fig:denma95} shows the observed data and fitted models for
Danish males born in 1895. The circles show the observed central death
rates and the area of each circle is proportional to the number of
person-years of exposure observed at each age. Figure~\ref{fig:denma95}
shows that there is much more information about the cohort's mortality
experience at age 80 than at age 100: by the oldest ages, the amount of
data has diminished dramatically, meaning that the data contain
relatively little information about the underlying mortality process at
oldest ages. The curves in Figure~\ref{fig:denma95} show the maximum
likelihood fit of each model from Table~\ref{tbl:hazfns}. Some models
apparently fit the observed data better than others: for example, the
Lynch-Brown model is able to bend at oldest ages to account for what may
be a deceleration in the observed central death rates, while the
Gompertz model appears to be forced to forgo fitting the small amount of
data at the highest ages in favor of closely fitting the majority of
deaths, which are concentrated at earlier ages.

Table~\ref{tbl:denma95} shows several quantitative summaries of the fits
shown in Figure~\ref{fig:denma95}. The models are ordered by the rank
they attain using the AIC, where rank 1 is the model that fits the data
the best. Table~\ref{tbl:denma95} also shows the log-likelihood, the sum
of squared errors in the estimated number of deaths at each age (SSE),
the SSE rank, and the difference between each model's AIC and the
minimum AIC (\(\Delta\)AIC). According to the SSE the fitted deaths from
Lynch-Brown model are closest to the observed deaths for this cohort.
However, the four-parameter Lynch-Brown model is very flexible, and the
SSE makes no adjustment to protect against overfitting.
Table~\ref{tbl:denma95} shows the AIC---which penalizes each model's
complexity to guard against overfitting---suggests that the Kannisto
model performs the best. Thus, although the Kannisto model is slightly
less accurate than the Lynch-Brown model, this is compensated for by the
fact that the Kannisto model uses only two parameters while the
Lynch-Brown model uses four. The AIC suggests that if we saw more data
from the same cohort, the fitted Kannisto model would be likely to fit
this unseen data more accurately than the fitted Lynch-Brown model.

\hypertarget{tbl:denma95}{}
\begin{longtable}[]{@{}ccccccc@{}}
\caption{\label{tbl:denma95}Measurements of model fit for the cohort of
Danish males born in 1895.}\tabularnewline
\toprule
\begin{minipage}[b]{0.14\columnwidth}\centering
~\strut
\end{minipage} & \begin{minipage}[b]{0.15\columnwidth}\centering
log-likelihood\strut
\end{minipage} & \begin{minipage}[b]{0.09\columnwidth}\centering
SSE\strut
\end{minipage} & \begin{minipage}[b]{0.10\columnwidth}\centering
SSE rank\strut
\end{minipage} & \begin{minipage}[b]{0.10\columnwidth}\centering
AIC\strut
\end{minipage} & \begin{minipage}[b]{0.10\columnwidth}\centering
AIC rank\strut
\end{minipage} & \begin{minipage}[b]{0.13\columnwidth}\centering
\(\Delta\) AIC\strut
\end{minipage}\tabularnewline
\midrule
\endfirsthead
\toprule
\begin{minipage}[b]{0.14\columnwidth}\centering
~\strut
\end{minipage} & \begin{minipage}[b]{0.15\columnwidth}\centering
log-likelihood\strut
\end{minipage} & \begin{minipage}[b]{0.09\columnwidth}\centering
SSE\strut
\end{minipage} & \begin{minipage}[b]{0.10\columnwidth}\centering
SSE rank\strut
\end{minipage} & \begin{minipage}[b]{0.10\columnwidth}\centering
AIC\strut
\end{minipage} & \begin{minipage}[b]{0.10\columnwidth}\centering
AIC rank\strut
\end{minipage} & \begin{minipage}[b]{0.13\columnwidth}\centering
\(\Delta\) AIC\strut
\end{minipage}\tabularnewline
\midrule
\endhead
\begin{minipage}[t]{0.14\columnwidth}\centering
Kannisto\strut
\end{minipage} & \begin{minipage}[t]{0.15\columnwidth}\centering
-29092.2\strut
\end{minipage} & \begin{minipage}[t]{0.09\columnwidth}\centering
7645.48\strut
\end{minipage} & \begin{minipage}[t]{0.10\columnwidth}\centering
6\strut
\end{minipage} & \begin{minipage}[t]{0.10\columnwidth}\centering
58188.43\strut
\end{minipage} & \begin{minipage}[t]{0.10\columnwidth}\centering
1\strut
\end{minipage} & \begin{minipage}[t]{0.13\columnwidth}\centering
0\strut
\end{minipage}\tabularnewline
\begin{minipage}[t]{0.14\columnwidth}\centering
Beard\strut
\end{minipage} & \begin{minipage}[t]{0.15\columnwidth}\centering
-29092.1\strut
\end{minipage} & \begin{minipage}[t]{0.09\columnwidth}\centering
7645.43\strut
\end{minipage} & \begin{minipage}[t]{0.10\columnwidth}\centering
3\strut
\end{minipage} & \begin{minipage}[t]{0.10\columnwidth}\centering
58190.12\strut
\end{minipage} & \begin{minipage}[t]{0.10\columnwidth}\centering
2\strut
\end{minipage} & \begin{minipage}[t]{0.13\columnwidth}\centering
1.7\strut
\end{minipage}\tabularnewline
\begin{minipage}[t]{0.14\columnwidth}\centering
Log-Quadratic\strut
\end{minipage} & \begin{minipage}[t]{0.15\columnwidth}\centering
-29092.1\strut
\end{minipage} & \begin{minipage}[t]{0.09\columnwidth}\centering
7645.45\strut
\end{minipage} & \begin{minipage}[t]{0.10\columnwidth}\centering
5\strut
\end{minipage} & \begin{minipage}[t]{0.10\columnwidth}\centering
58190.25\strut
\end{minipage} & \begin{minipage}[t]{0.10\columnwidth}\centering
3\strut
\end{minipage} & \begin{minipage}[t]{0.13\columnwidth}\centering
1.8\strut
\end{minipage}\tabularnewline
\begin{minipage}[t]{0.14\columnwidth}\centering
Gompertz\strut
\end{minipage} & \begin{minipage}[t]{0.15\columnwidth}\centering
-29093.4\strut
\end{minipage} & \begin{minipage}[t]{0.09\columnwidth}\centering
7645.81\strut
\end{minipage} & \begin{minipage}[t]{0.10\columnwidth}\centering
8\strut
\end{minipage} & \begin{minipage}[t]{0.10\columnwidth}\centering
58190.76\strut
\end{minipage} & \begin{minipage}[t]{0.10\columnwidth}\centering
4\strut
\end{minipage} & \begin{minipage}[t]{0.13\columnwidth}\centering
2.3\strut
\end{minipage}\tabularnewline
\begin{minipage}[t]{0.14\columnwidth}\centering
Lynch-Brown\strut
\end{minipage} & \begin{minipage}[t]{0.15\columnwidth}\centering
-29092\strut
\end{minipage} & \begin{minipage}[t]{0.09\columnwidth}\centering
7645.39\strut
\end{minipage} & \begin{minipage}[t]{0.10\columnwidth}\centering
1\strut
\end{minipage} & \begin{minipage}[t]{0.10\columnwidth}\centering
58192.06\strut
\end{minipage} & \begin{minipage}[t]{0.10\columnwidth}\centering
5\strut
\end{minipage} & \begin{minipage}[t]{0.13\columnwidth}\centering
3.6\strut
\end{minipage}\tabularnewline
\begin{minipage}[t]{0.14\columnwidth}\centering
Logistic\strut
\end{minipage} & \begin{minipage}[t]{0.15\columnwidth}\centering
-29092\strut
\end{minipage} & \begin{minipage}[t]{0.09\columnwidth}\centering
7645.42\strut
\end{minipage} & \begin{minipage}[t]{0.10\columnwidth}\centering
2\strut
\end{minipage} & \begin{minipage}[t]{0.10\columnwidth}\centering
58192.09\strut
\end{minipage} & \begin{minipage}[t]{0.10\columnwidth}\centering
6\strut
\end{minipage} & \begin{minipage}[t]{0.13\columnwidth}\centering
3.7\strut
\end{minipage}\tabularnewline
\begin{minipage}[t]{0.14\columnwidth}\centering
Perks\strut
\end{minipage} & \begin{minipage}[t]{0.15\columnwidth}\centering
-29092.1\strut
\end{minipage} & \begin{minipage}[t]{0.09\columnwidth}\centering
7645.44\strut
\end{minipage} & \begin{minipage}[t]{0.10\columnwidth}\centering
4\strut
\end{minipage} & \begin{minipage}[t]{0.10\columnwidth}\centering
58192.2\strut
\end{minipage} & \begin{minipage}[t]{0.10\columnwidth}\centering
7\strut
\end{minipage} & \begin{minipage}[t]{0.13\columnwidth}\centering
3.8\strut
\end{minipage}\tabularnewline
\begin{minipage}[t]{0.14\columnwidth}\centering
Makeham\strut
\end{minipage} & \begin{minipage}[t]{0.15\columnwidth}\centering
-29093.4\strut
\end{minipage} & \begin{minipage}[t]{0.09\columnwidth}\centering
7645.81\strut
\end{minipage} & \begin{minipage}[t]{0.10\columnwidth}\centering
7\strut
\end{minipage} & \begin{minipage}[t]{0.10\columnwidth}\centering
58192.76\strut
\end{minipage} & \begin{minipage}[t]{0.10\columnwidth}\centering
8\strut
\end{minipage} & \begin{minipage}[t]{0.13\columnwidth}\centering
4.3\strut
\end{minipage}\tabularnewline
\begin{minipage}[t]{0.14\columnwidth}\centering
Weibull\strut
\end{minipage} & \begin{minipage}[t]{0.15\columnwidth}\centering
-29154.8\strut
\end{minipage} & \begin{minipage}[t]{0.09\columnwidth}\centering
7656.04\strut
\end{minipage} & \begin{minipage}[t]{0.10\columnwidth}\centering
9\strut
\end{minipage} & \begin{minipage}[t]{0.10\columnwidth}\centering
58313.55\strut
\end{minipage} & \begin{minipage}[t]{0.10\columnwidth}\centering
9\strut
\end{minipage} & \begin{minipage}[t]{0.13\columnwidth}\centering
125\strut
\end{minipage}\tabularnewline
\bottomrule
\end{longtable}

\hypertarget{fig:denma95}{%
\begin{figure}
\centering
\includegraphics{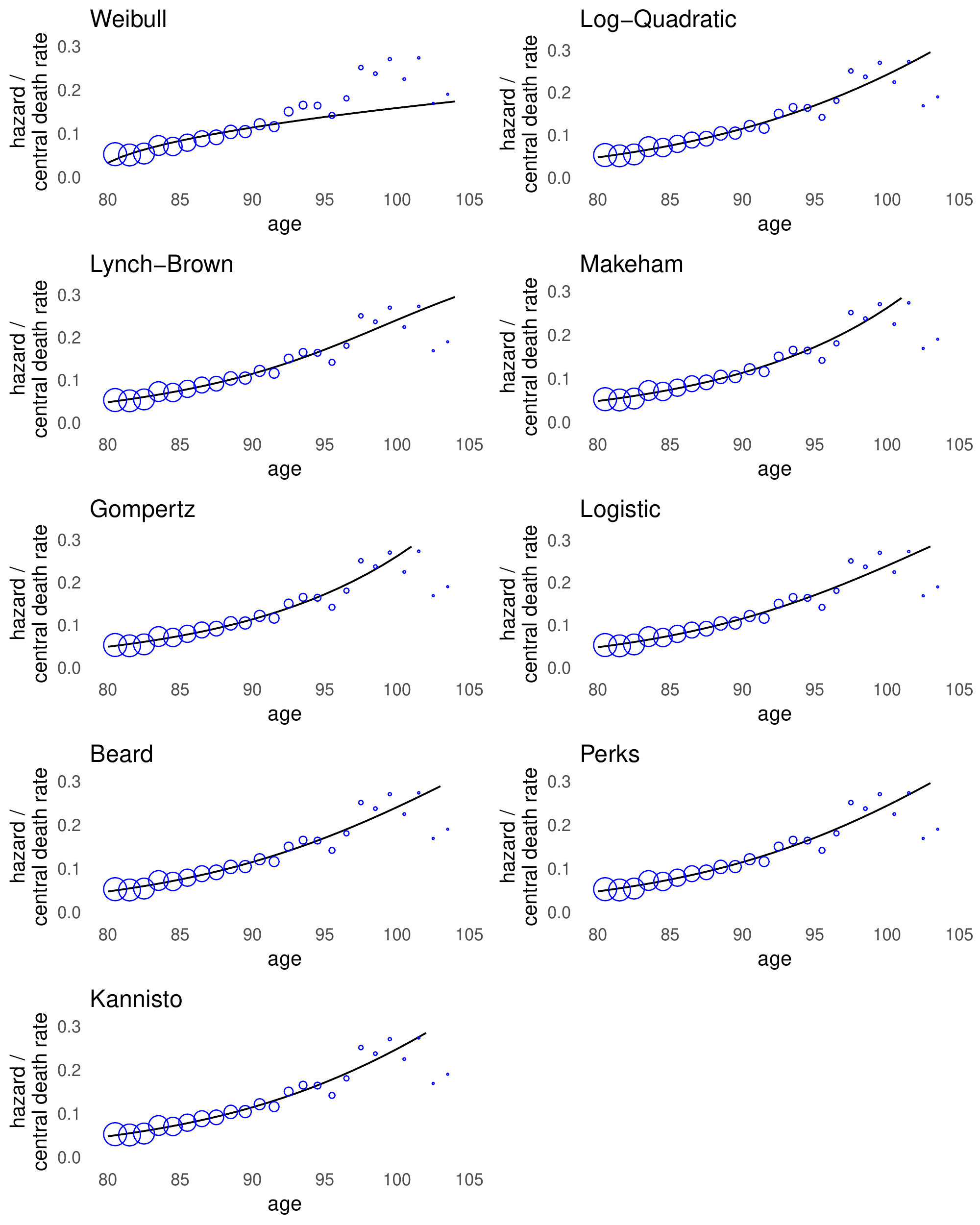}
\caption{Models fit to the cohort data for Danish males born in 1895.
The circles show the observed central death rates, the area of each blue
circle is proportional to the estimated person-years of exposure at each
age, and the curves show the maximum likelihood fit of each model.
Quantitative summaries of goodness of fit are shown in
Table~\ref{tbl:denma95}. The AIC suggests that the Kannisto model best
fits this cohort.}\label{fig:denma95}
\end{figure}
}

\hypertarget{sec:results-all}{%
\subsection{Results for all cohorts}\label{sec:results-all}}

Figure~\ref{fig:deltaaicboxplot} shows boxplots that summarize the
distribution of \(\Delta\text{AIC}\) across all 360 cohorts, by sex.
Note that these results essentially weight the mortality experience of
each country by the number of cohorts that it contributes to the sample
(Table~\ref{tbl:kt-data}). Several observations emerge from this figure:
first, according to the median \(\Delta\text{AIC}\) value, no single
model clearly performs the best; instead, several of the models have
very similar performance: for example, across both sexes, the
Log-Quadratic, Beard, and Kannisto models have the lowest median
\(\Delta\text{AIC}\) values, and those median \(\Delta\text{AIC}\)
values are all very close to one another. Second,
Figure~\ref{fig:deltaaicboxplot} shows that there is much more variation
in \(\Delta\text{AIC}\) for some models than there is for others; for
example, the interquartile range in \(\Delta\text{AIC}\) across female
cohorts is 2 for the Perks model, while it is 15 for the Kannisto model.
The Weibull model consistently performed poorly, so we do not discuss it
further.

\hypertarget{fig:deltaaicboxplot}{%
\begin{figure}
\centering
\includegraphics{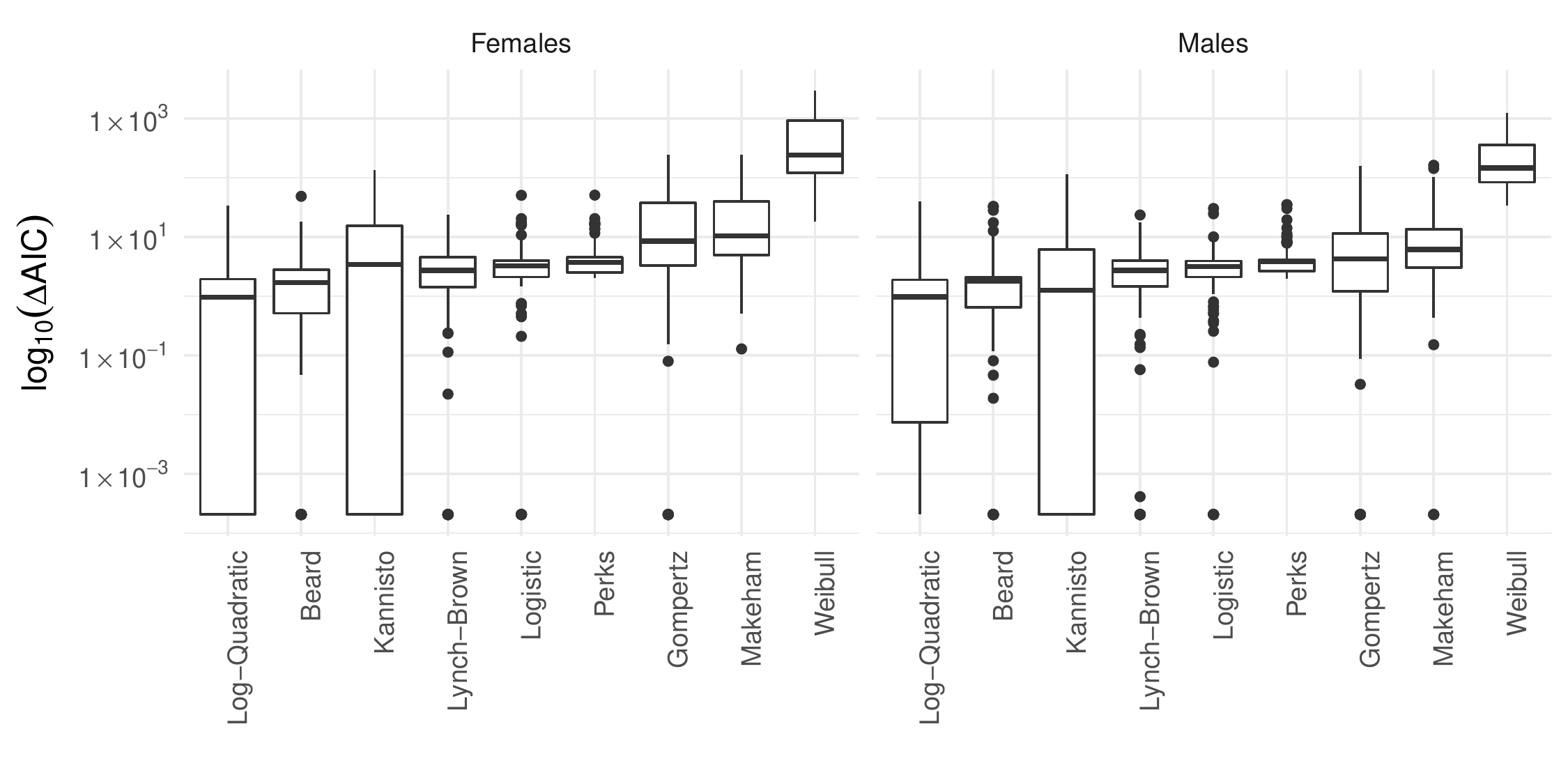}
\caption{Boxplots summarizing the distributions of \(\Delta\text{AIC}\)
across all of the cohorts in the sample (note that the y axis is on a
log scale). For each box, the horizontal line shows the median, the
edges of the box show the interquartile range, and the whiskers extend
to the largest and smallest values within 1.5 times the interquartile
range; more extreme points are plotted separately. The Log-Quadratic,
Beard, and Kannisto models appear to perform the
best.}\label{fig:deltaaicboxplot}
\end{figure}
}

The variation in \(\Delta\text{AIC}\) (Figure~\ref{fig:deltaaicboxplot})
reveals that there is more than one factor to consider when evaluating
the fit of each model. It is desirable to find a model that (1) often
fits the data well and (2) rarely fits the data poorly. Burnham and
Anderson (2003) propose rules of thumb that can help interpret
\(\Delta\text{AIC}\) with these two goals in mind: models with
`substantial support' from the data have \(\Delta\text{AIC} \leq 2\)
while models with `essentially no support' from the data have
\(\Delta\text{AIC} > 10\) (Burnham and Anderson 2003, pg. 72).
Figure~\ref{fig:deltaaicgoodbad} uses these rules of thumb to compare
the fraction of cohorts for which each model fits the data well (x axis)
and the fraction of cohorts for which each model fits the data poorly (y
axis). Several observations can be made about
Figure~\ref{fig:deltaaicgoodbad}. First, according to
\(\Delta\text{AIC}\) the Log-Quadratic model appears to come closest to
satisfying the two goals: it often fits the data well
(\(\Delta\text{AIC} \leq 2\) for more than three quarters of cohorts)
and it rarely fits the data poorly (\(\Delta\text{AIC} > 10\) for fewer
than one tenth of cohorts); the Beard model performs almost as well as
Log-Quadratic. Second, several models that do not fit the data extremely
well also avoid fitting the data extremely badly. In particular, the
most flexible models--- Perks, Logistic, and Lynch-Brown---do not often
fit the data the best (all have \(\Delta\text{AIC} \leq 2\) for less
than half of the cohorts), but these four-parameter models also rarely
fit the data very poorly (all have \(\Delta\text{AIC} > 10\) for less
than one tenth of the cohorts). Finally,
Figure~\ref{fig:deltaaicgoodbad} reveals that most of the patterns of
model fit are very similar for male and for female cohorts. There are
three notable exceptions: the Kannisto, Gompertz, and Makeham models all
show bigger differences between the fits for males and females than the
other models do. For all three of these models, the point on
Figure~\ref{fig:deltaaicgoodbad} summarizing fits to male cohorts is
below and to the right of the point summarizing fits to female cohorts;
thus, the Gompertz, Makeham, and Kannisto models systematically do a
better job of capturing the patterns in male mortality than female
mortality.

\hypertarget{fig:deltaaicgoodbad}{%
\begin{figure}
\centering
\includegraphics{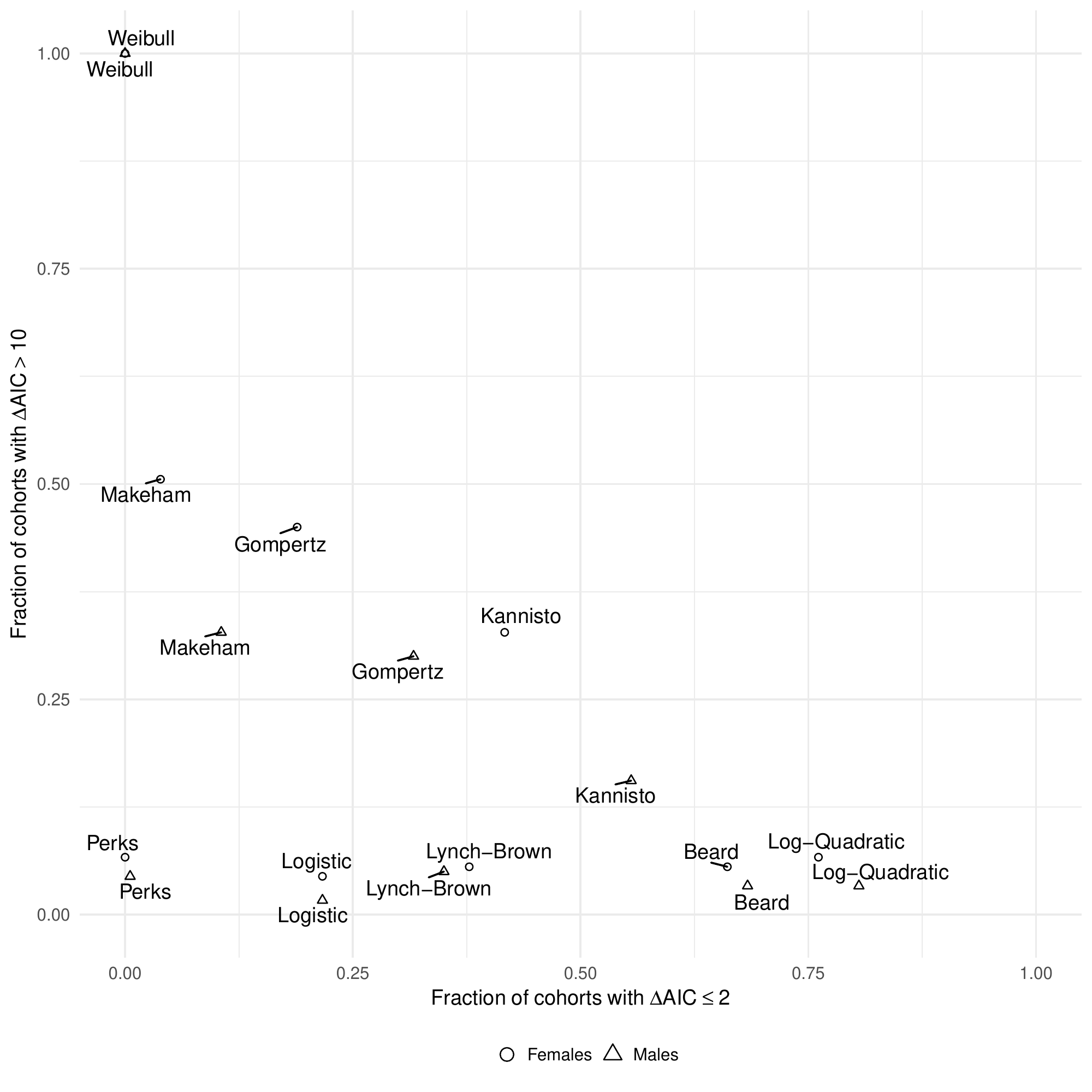}
\caption{Fraction of cohorts for which the fit of each model had
substantial support (\(\Delta\text{AIC} <= 2\), x axis) compared to the
fraction of cohorts for which the fit of each model had essentially no
support (\(\Delta\text{AIC} > 10\), y axis). Models in the lower-right
corner consistently fit the cohorts in the sample well; models in the
lower-left corner are neither very good nor very bad; models in the
upper-left corner consistently fit the cohorts in the sample
poorly.}\label{fig:deltaaicgoodbad}
\end{figure}
}

Next, we investigate the possibility that there are pairs or groups of
models that tend to perform relatively well or to perform relatively
poorly for the same cohorts. Figure~\ref{fig:deltaaic-dend} shows a
dendrogram that describes the similarity between patterns of
\(\Delta\text{AIC}\) for each model across all 360 cohorts\footnote{The
  distance metric used in Appendix Figure~\ref{fig:deltaaic-dend} is
  group average difference in \(\Delta\text{AIC}\) values. Friedman et
  al. (2009) has more details on hierarchical clustering.}.
Figure~\ref{fig:deltaaic-dend} suggests that a natural grouping can be
obtained by picking a dissimilarity of about 200, which results in three
groups of models: the first group contains the Log-Quadratic, Perks,
Beard, Logistic, and Lynch-Brown models; the second group contains the
Kannisto model; and the third group contains the Gompertz and Makeham
models. Figure~\ref{fig:deltaaicboxplot} and
Figure~\ref{fig:deltaaicgoodbad} show that models from the first
group---which consists of functional forms that can bend to accommodate
decelerating death rates at older ages---tend to fit the data well: they
rarely have \(\Delta\text{AIC} \geq 10\), and the Log-Quadratic and
Beard models have the highest fraction of cohorts with
\(\Delta\text{AIC} < 2\). In the second group, the Kannisto model's
relative performance is more ambiguous because there is a large amount
of variance in the quality of its fits; for some cohorts, Kannisto
performs as well as models from the first group, but for other cohorts,
Kannisto's performance is more similar to the Gompertz/Makeham models.
Finally, in the third group, the Gompertz and Makeham models clearly
tend to perform worse than the more flexible models in the first group,
suggesting that the steady exponential increase in death rates observed
at middle adult ages tends not to continue into advanced ages.

\hypertarget{fig:deltaaic-dend}{%
\begin{figure}
\centering
\includegraphics{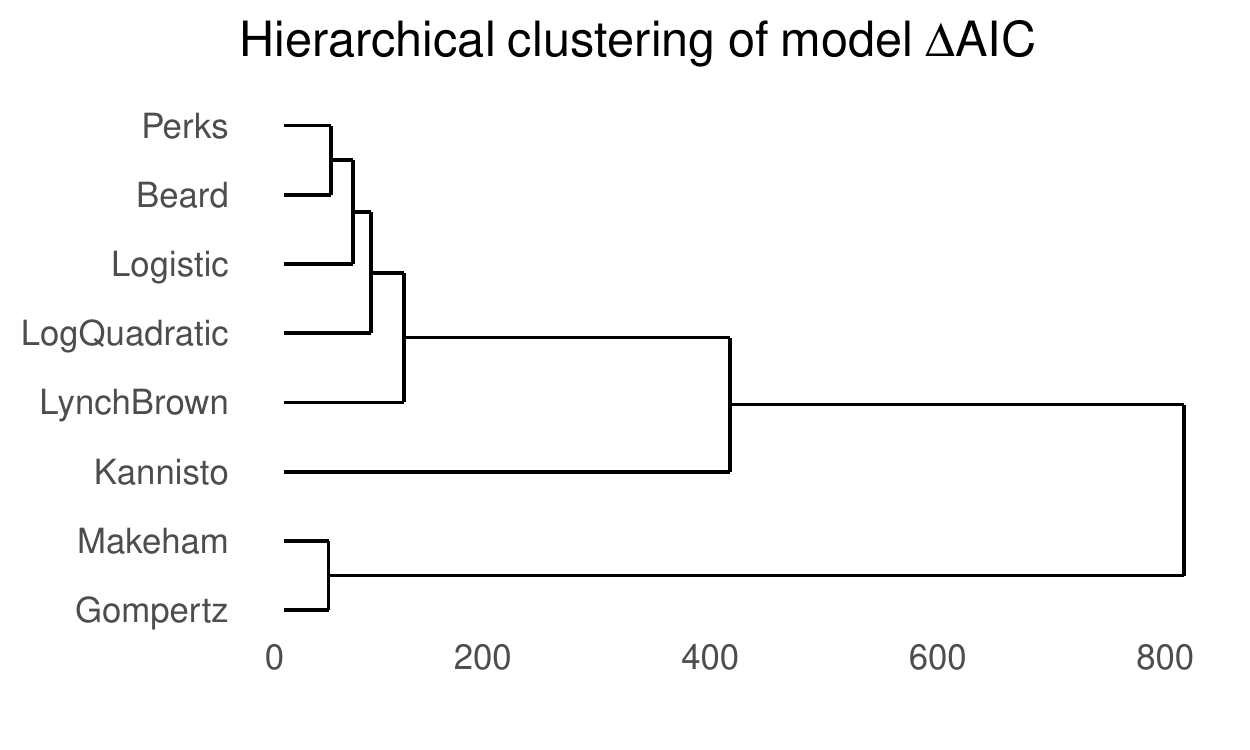}
\caption{Dendrogram showing the relationship between model
\(\Delta\text{AIC}\) across all of the cohorts in our sample. The
Weibull model was omitted, since it provides uniformly poor fits.\\
Each node in the tree corresponds to a clustering of models into two
groups, one given by the tree corresponding to the node's top child and
one to the tree corresponding to the node's bottom child. The position
of the node along the x axis is proportional to the dissimilarity in
between the two clusters; this dissimilarity is calculated as the
Euclidean distance between the two cluster averages. The entire
dendrogram shows a nested sequence of clusterings, called a
\emph{hierarchical clustering}; picking any dissimilarity value on the x
axis produces a clustering.}\label{fig:deltaaic-dend}
\end{figure}
}

\hypertarget{sec:results-country}{%
\subsection{Results by country}\label{sec:results-country}}

The results in Section~\ref{sec:results-all} summarize model fits across
the entire sample, which consists of cohorts from many different
countries. We now turn to an analysis of model fits within each country,
focusing on the five countries where more than 10 cohorts are observed
for each sex: Denmark, France, Italy, the Netherlands, and Sweden. The
remaining five countries---Belgium, West Germany, Japan, Scotland, and
Switzerland---had 10 or fewer cohorts each; to avoid the risk of making
misleading inferences from so few cohorts, we omit them from this
country-specific analysis.

Figure~\ref{fig:deltaaicgoodbadcountry} again uses the rules of thumb to
show the fraction of cohorts for which each model has considerable
support (x axis) and the fraction of cohorts for which each model has
essentially no support (y axis) within each country and sex.\footnote{The
  figure omits the Weibull model, which consistently fit poorly across
  all countries (Figure~\ref{fig:deltaaicgoodbad}).} The results suggest
that (1) there are big differences in how the models fit cohorts from
different countries, but (2) within each country there tend not to be
big differences in model fits by sex. The performance of the Gompertz
and the Makeham models distinguishes countries from one another: in
France and Italy, the Gompertz and Makeham models clearly perform
poorly, and more flexible models fit well. For Denmark, the Gompertz and
Makeham models do not rank at the top, but they rarely predict death
rates very poorly (\(\Delta\text{AIC} > 10\) in few cohorts). Sweden and
the Netherlands are intermediate cases where Gompertz and Makeham
predict poorly under half of the time.

\hypertarget{fig:deltaaicgoodbadcountry}{%
\begin{figure}
\centering
\includegraphics[width=\textwidth,height=1.1\textwidth]{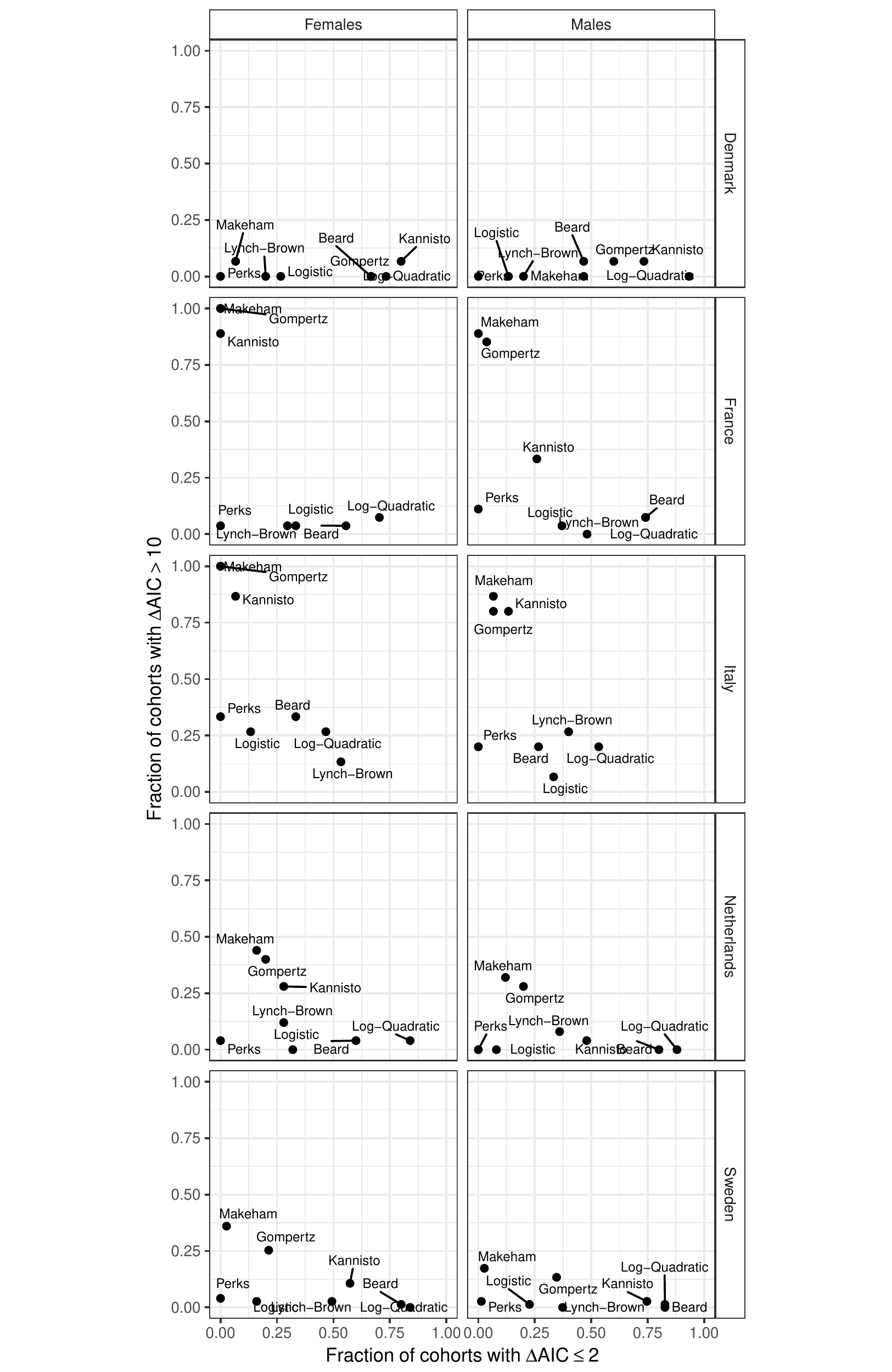}
\caption{Fraction of cohorts for which the fit of each model had
substantial support (\(\Delta\text{AIC} <= 2\), x axis) compared to the
fraction of cohorts for which the fit of each model had essentially no
support (\(\Delta\text{AIC} > 10\), y axis) for each sex and for each
country with at least ten cohorts of data. Models in the lower-right
corner consistently fit the cohorts in the sample well; models in the
lower-left corner are neither very good nor very bad; models in the
upper-left corner consistently fit the cohorts in the sample
poorly.}\label{fig:deltaaicgoodbadcountry}
\end{figure}
}

\hypertarget{summary-and-discussion}{%
\subsection{Summary and Discussion}\label{summary-and-discussion}}

The Gompertz model provides a remarkably good description of death rates
at young and middle adult ages, and researchers have long debated
whether or not Gomepertzian increases in death rates continue
indefinitely with age (\emph{e.g.} Bebbington et al. 2014; Olshansky and
Carnes 1997; A. Thatcher et al. 1998; Vaupel et al. 1998). A key finding
of our analysis is that flexible, non-Gompertzian models tend to fit
old-age death rates better than Gompertz models in our sample of 360
high-quality cohorts from the Kannisto-Thatcher database\footnote{Appendix
  \ref{sec:ap-results} shows that this conclusion is robust to the
  choice of AIC, BIC, or cross-validaton as the principled model fit
  technique.}. This finding supports the idea that death rates tend to
decelerate at oldest ages. Critically, we reached this conclusion by
using a principled approach to model selection which accounted for the
fact that there are very small numbers of deaths at advanced ages.

Our findings are consistent with several previous studies that have
uncovered evidence of non-Gompertzian patterns in old-age death rates
using other methods or data sources (Horiuchi and Wilmoth 1998; A.
Thatcher et al. 1998; Vaupel et al. 1998; Yi and Vaupel 2003). However,
recent analyses have also found support for Gompertz-type models of
old-age mortality in some cases. Bebbington et al. (2014) used the Human
Mortality Database to find that old-age mortality patterns for most
countries do not follow a Gompertz-type trajectory; however, Bebbington
et al. (2014) identified three notable exceptions where the Gompertz
model fit well: Australia, Canada, and the United States. This raises
the possibility that there may be a group of countries for which the
Gompertz model provides a good description of old-age mortality. So far,
this hypothesis has mixed support: in the United States, L. A. Gavrilov
and Gavrilova (2011) and N. S. Gavrilova and Gavrilov (2015) analyzed
the US Social Security Death Master File and found additional evidence
supporting Gompertzian old-age mortality in the US. In Canada, however,
Ouellette (2016) finds non-Gompertz mortality trajectories in old-age
mortality among French Canadians in Quebec. Further, recent research in
actuarial modeling suggests that non-Gompertzian dynamics may appear in
Australian old-age mortality (J. S. H. Li et al. 2011). Thus, recent
studies support the suggestion that US old-age mortality dynamics may be
Gompertzian, but there are mixed results for Australia and Canada. Our
study cannot directly address the debate in these three countries, since
the Kannisto-Thatcher data from Australia, Canada, and the United States
were not of the highest quality, and so they were excluded from this
analysis.

Although we are able to conclude that flexible models from the first
group tend to perform better than the Gompertz/Makeham models in the
third group, our results do not support the idea that any particular one
of these nine models universally describes old-age mortality. The
Log-Quadratic model most frequently provided accurate fits, but even the
Log-Quadratic only had substantial support from the data
(\(\Delta\text{AIC} \leq 2\)) in about three quarters of the cohorts
studied; thus, even the Log-Quadratic model cannot be said to
comprehensively capture all of the patterns in our sample.

We also found that patterns of model fit differed by country. Models
from the Gompertz group did not reliably fit well
(\(\Delta\text{AIC} < 2\)) in any of the countries; however, countries
can be differentiated according to the fraction of cohorts in which the
Gompertz group models performed poorly (\(\Delta\text{AIC} > 10\)).
France and Italy stand out because in these two countries models from
the Gompertz group were clearly inadequate: they did not often fit well,
and frequently fit poorly\footnote{This finding that France and Italy do
  not fit Gompertzian old-age mortality patterns is confirmed even when
  other methods of assessing model fit, based on different statistical
  theories, are considered (Appendix \ref{sec:ap-results}).}. Sweden and
the Netherlands are intermediate cases in which models from the Gompertz
group tended not to provide the best fit, but the Gompertz group models
only performed poorly for about half of the cohorts. Finally, in Denmark
models from the Gompertz group tended not to provide the best fit, but
performed the best among the countries considered here. (In Denmark,
none of the models consistently performed poorly.)

There are several possible explanations for these differing patterns by
country, and these explanations are not mutually exclusive. First, it is
possible that relatively large cohort sizes in Italy and France provide
enough statistical power to detect deviations from Gompertzian dynamics,
while smaller cohort sizes in Denmark, Sweden, and the Netherlands make
this more difficult, even if underlying mortality dynamics in the
smaller countries are non-Gompertzian. Although this possibility seems
compelling, initial investigations revealed that artificially sampling
successively smaller cohort sizes from the French data did not
automatically lead to better performance by the Gompertz models, but
instead tended to eventually produce better performance by the Kannisto
model (Appendix \ref{sec:samplesize}). Thus, although sample size is
clearly a factor in our ability to detect deviations from
non-Gompertzian mortality dynamics, it seems unlikely that it is the
only explanation for the country-specific patterns reported here.

Second, it is possible that country-specific differences in the timing
and rate of change of death rates at oldest ages leads to different
patterns of fit by country. This view is supported by the fact that the
country groupings suggested by our analysis are consistent with previous
studies of levels and changes in adult mortality in Europe: other
researchers have observed a contrast between Southern European countries
like France and Italy, which showed continual improvement in adult death
rates since the middle of the twentieth century, and some Northern
European countries, including Denmark and the Netherlands, whose adult
mortality gains have been less even, with periods of relative stagnation
(Luy et al. 2011; Meslé and Vallin 2006; R. Rau et al. 2008). Staetsky
(2009) argues that these contrasting patterns in female mortality change
can be explained by differences in the timing and intensity of smoking,
with women in Denmark and the Netherlands starting to smoke in earlier
time periods than their counterparts in Italy and France. Interestingly,
Staetsky (2009) points out that the timing of smoking in Denmark and the
Netherlands closely follows the trend in the United States, a country
where Bebbington et al. (2014) and L. A. Gavrilov and Gavrilova (2011)
found support for Gompertz patterns of old-age mortality. Thus, our
country-specific results suggest that period effects may have altered
the shape of cohort mortality, so that cohorts that experienced
continual improvements (France and Italy) show the strongest evidence of
non-Gompertz death rates.

Third, these country-specific differences in model fits could be the
result of more fundamental differences in the mortality regime of each
country. These differences could be produced by country-specific
differences in genetic or cultural factors--such as patterns of diet,
physical activity, and social support--that are known to be important
predictors of adult mortality and are thus likely to be moderators of
old-age mortality (Berkman and Syme 1979; Byberg et al. 2009; Gjonca et
al. 2000; Haveman-Nies et al. 2003; Vaupel et al. 2003).

Finally, we examined patterns of model fit by sex. Across all cohorts,
most models showed similar performance for males and for females. The
exceptions were the Gompertz, Makeham, and Kannisto models, which
systematically tended to fit male cohorts better than female cohorts
(Figure~\ref{fig:deltaaicgoodbad}). More generally, patterns of model
fit varied more across countries than across sexes; this finding was
somewhat surprising, since adult mortality levels are often found to be
more similar among cohorts of the same sex from different countries, as
opposed to cohorts of different sexes from the same country (e.g., Fig.
3.4 in Luy et al. 2011). We consider understanding this finding an
important topic for future research.

\hypertarget{sec:conclusion}{%
\section{Conclusion}\label{sec:conclusion}}

These results lead us to offer some suggestions for researchers who wish
to model old-age mortality in the future. First, researchers who develop
and fit mortality models should use a principled model selection
technique like the AIC to assess model fit. Second, if a single model is
to be used to model old-age mortality hazards, our results suggest that
the Log-Quadratic model is a reasonable choice. Other models can also
perform well, though our results showed that the Kannisto model---which
is often used in practice---should be chosen with caution. Finally, our
results suggest that researchers should be cautious about designing
mortality studies that exclusively use data from only one country,
particularly if cohort sizes from the country tend to be small.

This study has several limitations and the results suggest several
directions for future research. Any study of old-age mortality is
limited by the quantity and quality of data available. We restricted our
analysis to cohorts in the Kannisto-Thatcher database that have been
identified as having the highest quality data, but it is possible that
some inaccuracies remain. Severe data inaccuracies could affect model
fits and thus our conclusions.

More generally, like many analyses of old-age mortality, we are
restricted to analyzing groups of people that have been defined by
political boundaries (i.e.~countries). This approach is reasonable
because people who live in the same country are governed by the same
national policy regime; further, people who live in the same country may
tend to be genetically similar and may tend to live in similar cultural
environments. However, there can be a great deal of variation within a
single country. Most models of mortality---even those models that
explicitly account for heterogeneity---are most appropriate for one
\emph{coherent} population whose members face some continuous
distribution of hazards. Countries that are big and diverse may well be
better described as finite mixtures of coherent populations. Thus, it
may be productive to try to construct datasets for cohorts whose members
are most likely to share a common characteristic mortality experience,
possibly with individual variation.

Our design focused on analyzing high-quality information on deaths and
exposure for entire cohorts. Other designs based on collecting more
detailed information from samples of people---such as longitudinal
health surveys that link respondents to death certificate data---should
also be used to assess the empirical support for different theories of
old-age mortality. These designs have many advantages, as well as a
different set of methodological and data quality challenges; thus,
longitudinal surveys can be seen as an important complement to the
analysis here.

Finally, we hope this analysis can be a helpful starting point for
building more complex models to understand and to forecast old-age
mortality. Future work could investigate how to combine these
theoretical models into predictive ensembles; researchers could also
build a hierarchical model that describes how cohorts in the sample are
related to one another. Either or both of these approaches could be used
to stochastically forecast future old-age death rates.

\newpage

\hypertarget{references}{%
\section{References}\label{references}}

\hypertarget{refs}{}
\leavevmode\hypertarget{ref-abramowitz_handbook_1964}{}%
Abramowitz, M., \& Stegun, I. A. (1964). \emph{Handbook of mathematical
functions: With formulas, graphs, and mathematical tables} (Vol. 55).
Courier Corporation.

\leavevmode\hypertarget{ref-akaike_new_1974}{}%
Akaike, H. (1974). A new look at the statistical model identification.
\emph{IEEE transactions on automatic control}, \emph{19}(6), 716--723.

\leavevmode\hypertarget{ref-beard_appendix_1959}{}%
Beard, R. E. (1959). Appendix: Note on Some Mathematical Mortality
Models, 302--311.

\leavevmode\hypertarget{ref-beard_aspects_1971}{}%
Beard, R. E. (1971). Some aspects of theories of mortality, cause of
death analysis, forecasting and stochastic processes. \emph{Biological
aspects of demography}, 57--68.

\leavevmode\hypertarget{ref-bebbington_gompertz_2014}{}%
Bebbington, M., Green, R., Lai, C.-D., \& Zitikis, R. (2014). Beyond the
Gompertz law: Exploring the late-life mortality deceleration phenomenon.
\emph{Scandinavian Actuarial Journal}, \emph{2014}(3), 189--207.

\leavevmode\hypertarget{ref-berkman_social_1979}{}%
Berkman, L. F., \& Syme, S. L. (1979). Social networks, host resistance,
and mortality: A nine-year follow-up study of Alameda County residents.
\emph{American Journal of Epidemiology}, \emph{109}(2), 186--204.

\leavevmode\hypertarget{ref-black_methuselah_2017}{}%
Black, D. A., Hsu, Y.-C., Sanders, S. G., Schofield, L. S., \& Taylor,
L. J. (2017). \emph{The Methuselah Effect: The Pernicious Impact of
Unreported Deaths on Old Age Mortality Estimates}. National Bureau of
Economic Research.

\leavevmode\hypertarget{ref-bongaarts_longrange_2005}{}%
Bongaarts, J. (2005). Long-range trends in adult mortality: Models and
projection methods. \emph{Demography}, \emph{42}(1), 23--49.

\leavevmode\hypertarget{ref-burnham_model_2003}{}%
Burnham, K. P., \& Anderson, D. (2003). Model selection and multi-model
inference: A practical information-theoretic approach. \emph{Springer}.

\leavevmode\hypertarget{ref-burnham_multimodel_2004}{}%
Burnham, K. P., \& Anderson, D. R. (2004). Multimodel Inference:
Understanding AIC and BIC in Model Selection. \emph{Sociological Methods
\& Research}, \emph{33}(2), 261--304.

\leavevmode\hypertarget{ref-byberg_total_2009}{}%
Byberg, L., Melhus, H., Gedeborg, R., Sundström, J., Ahlbom, A.,
Zethelius, B., et al. (2009). Total mortality after changes in leisure
time physical activity in 50 year old men: 35 year follow-up of
population based cohort. \emph{BMJ}, \emph{338}, b688.

\leavevmode\hypertarget{ref-chiang_stochastic_1960}{}%
Chiang, C. L. (1960). A stochastic study of the life table and its
applications: I. Probability distributions of the biometric functions.
\emph{Biometrics}, \emph{16}(4), 618--635.

\leavevmode\hypertarget{ref-claeskens_model_2008}{}%
Claeskens, G., \& Hjort, N. L. (2008). \emph{Model selection and model
averaging} (Vol. 330). Cambridge University Press Cambridge.

\leavevmode\hypertarget{ref-coale_defects_1990}{}%
Coale, A. J., \& Kisker, E. E. (1990). Defects in data on old-age
mortality in the United States: New procedures for calculating mortality
schedules and life tables at the highest ages.

\leavevmode\hypertarget{ref-daroczi_pander_2015}{}%
Daróczi, G., \& Tsegelskyi, R. (2015). \emph{Pander: An R Pandoc Writer.
R package version 0.6.0}.

\leavevmode\hypertarget{ref-dong_evidence_2016}{}%
Dong, X., Milholland, B., \& Vijg, J. (2016). Evidence for a limit to
human lifespan. \emph{Nature}, \emph{538}(7624), 257--259.

\leavevmode\hypertarget{ref-efron_computer_2016}{}%
Efron, B., \& Hastie, T. (2016). \emph{Computer Age Statistical
Inference} (Vol. 5). Cambridge University Press.

\leavevmode\hypertarget{ref-friedman_elements_2009}{}%
Friedman, J., Hastie, T., \& Tibshirani, R. (2009). The Elements of
Statistical Learning: Data Mining, Inference, and Prediction.
\emph{Springer Series in Statistics}.

\leavevmode\hypertarget{ref-gavrilov_biology_1991}{}%
Gavrilov, L. A., \& Gavrilova, N. S. (1991). \emph{The biology of life
span: A quantitative approach}. Chur {[}Switzerland{]}; New York:
Harwood Academic Publishers.

\leavevmode\hypertarget{ref-gavrilov_mortality_2011}{}%
Gavrilov, L. A., \& Gavrilova, N. S. (2011). Mortality measurement at
advanced ages: A study of the Social Security Administration Death
Master File. \emph{North American actuarial journal: NAAJ},
\emph{15}(3), 432.

\leavevmode\hypertarget{ref-gavrilova_biodemography_2015}{}%
Gavrilova, N. S., \& Gavrilov, L. A. (2015). Biodemography of Old-Age
Mortality in Humans and Rodents. \emph{The Journals of Gerontology:
Series A}, \emph{70}(1), 1--9.

\leavevmode\hypertarget{ref-gerland_world_2014}{}%
Gerland, P., Raftery, A. E., Ševčíková, H., Li, N., Gu, D., Spoorenberg,
T., et al. (2014). World population stabilization unlikely this century.
\emph{Science}, \emph{346}(6206), 234--237.

\leavevmode\hypertarget{ref-gjonca_oldage_2000}{}%
Gjonca, A., Maier, H., \& Brockmann, H. (2000). Old-Age Mortality in
Germany prior to and after Reunification. \emph{Demographic Research},
\emph{3}.

\leavevmode\hypertarget{ref-gompertz_nature_1825}{}%
Gompertz, B. (1825). On the Nature of the Function Expressive of the Law
of Human Mortality, and on a New Mode of Determining the Value of Life
Contingencies. \emph{Philosophical Transactions of the Royal Society of
London}, \emph{115}, 513--583.

\leavevmode\hypertarget{ref-haveman-nies_dietary_2003}{}%
Haveman-Nies, A., Groot, D., C.p.g.m, L., Staveren, V., \& A, W. (2003).
Dietary quality, lifestyle factors and healthy ageing in Europe: The
SENECA study. \emph{Age and Ageing}, \emph{32}(4), 427--434.

\leavevmode\hypertarget{ref-himes_relational_1994}{}%
Himes, C., Preston, S., \& Condran, G. (1994). A relational model of
mortality at older ages in low mortality countries. \emph{Population
Studies}, \emph{48}(2), 269--291.

\leavevmode\hypertarget{ref-hoeting_bayesian_1999}{}%
Hoeting, J. A., Madigan, D., Raftery, A. E., \& Volinsky, C. T. (1999).
Bayesian model averaging: A tutorial. \emph{Statistical science},
382--401.

\leavevmode\hypertarget{ref-horiuchi_interspecies_2003}{}%
Horiuchi, S. (2003). Interspecies differences in the life span
distribution: Humans versus invertebrates. \emph{Population and
Development Review}, \emph{29}, 127--151.

\leavevmode\hypertarget{ref-horiuchi_deceleration_1998}{}%
Horiuchi, S., \& Wilmoth, J. R. (1998). Deceleration in the age pattern
of mortality at olderages. \emph{Demography}, \emph{35}(4), 391--412.

\leavevmode\hypertarget{ref-jdanov_kannisto_2008}{}%
Jdanov, D. A., Jasilionis, D., Soroko, E. L., Rau, R., \& Vaupel, J. W.
(2008). Beyond the Kannisto-Thatcher Database on Old-Age Mortality: An
assessment of data quality at advanced ages. \emph{MPIDR Working
Papers}.

\leavevmode\hypertarget{ref-kannisto_reductions_1994}{}%
Kannisto, V., Lauritsen, J., Thatcher, A. R., \& Vaupel, J. W. (1994).
Reductions in mortality at advanced ages: Several decades of evidence
from 27 countries. \emph{Population and Development Review}, 793--810.

\leavevmode\hypertarget{ref-kass_bayes_1995}{}%
Kass, R. E., \& Raftery, A. E. (1995). Bayes Factors. \emph{Journal of
the American Statistical Association}, \emph{90}(430), 773--795.

\leavevmode\hypertarget{ref-kinsella_global_2005}{}%
Kinsella, K., Phillips, D., \& Bureau, P. R. (2005). \emph{Global aging:
The challenge of success}. Population Reference Bureau.

\leavevmode\hypertarget{ref-lebras_lois_1976}{}%
Le Bras, H. (1976). Lois de mortalité et age limite. \emph{Population
(French Edition)}, 655--692.

\leavevmode\hypertarget{ref-lenart_questionable_2017}{}%
Lenart, A., \& Vaupel, J. W. (2017). Questionable evidence for a limit
to human lifespan. \emph{Nature}, \emph{546}(7660), E13.

\leavevmode\hypertarget{ref-li_modeling_2011}{}%
Li, J. S. H., Ng, A. C. Y., \& Chan, W. S. (2011). Modeling old-age
mortality risk for the populations of Australia and New Zealand: An
extreme value approach. \emph{Mathematics and Computers in Simulation},
\emph{81}(7), 1325--1333.

\leavevmode\hypertarget{ref-luy_adult_2011}{}%
Luy, M., Wegner, C., \& Lutz, W. (2011). Adult mortality in Europe. In
\emph{International handbook of adult mortality} (pp. 49--81). Springer.

\leavevmode\hypertarget{ref-lynch_reconsidering_2001}{}%
Lynch, S. M., \& Brown, J. S. (2001). Reconsidering mortality
compression and deceleration: An alternative model of mortality rates.
\emph{Demography}, \emph{38}(1), 79--95.

\leavevmode\hypertarget{ref-lynch_blackwhite_2003}{}%
Lynch, S. M., Brown, J. S., \& Harmsen, K. G. (2003). Black-White
Differences in Mortality Compression and Deceleration and the Mortality
Crossover Reconsidered. \emph{Research on Aging}, \emph{25}(5),
456--483.

\leavevmode\hypertarget{ref-makeham_law_1860}{}%
Makeham, W. M. (1860). On the law of mortality and the construction of
annuity tables. \emph{The Assurance Magazine, and Journal of the
Institute of Actuaries}, 301--310.

\leavevmode\hypertarget{ref-manton_methods_1981}{}%
Manton, K. G., Stallard, E., \& Vaupel, J. W. (1981). Methods for
comparing the mortality experience of heterogeneous populations.
\emph{Demography}, \emph{18}(3), 389--410.

\leavevmode\hypertarget{ref-manton_alternative_1986}{}%
Manton, K. G., Stallard, E., \& Vaupel, J. W. (1986). Alternative models
for the heterogeneity of mortality risks among the aged. \emph{Journal
of the American Statistical Association}, \emph{81}(395), 635--644.

\leavevmode\hypertarget{ref-martin_demography_1994}{}%
Martin, L., \& Preston, S. (1994). \emph{Demography of aging}. National
Academies Press.

\leavevmode\hypertarget{ref-mesle_diverging_2006}{}%
Meslé, F., \& Vallin, J. (2006). Diverging Trends in Female Old-Age
Mortality: The United States and the Netherlands versus France and
Japan. \emph{Population and Development Review}, \emph{32}(1), 123--145.

\leavevmode\hypertarget{ref-mpidr_kannistothatcher_2014}{}%
MPIDR. (2014). Kannisto-Thatcher Database on Old-Age Mortality.
http://www.demogr.mpg.de/databases/ktdb/introduction.htm.

\leavevmode\hypertarget{ref-nocedal_numerical_1999}{}%
Nocedal, J., \& Wright, S. (1999). \emph{Numerical optimization}.
Springer Verlag.

\leavevmode\hypertarget{ref-olshansky_ever_1997}{}%
Olshansky, S. J., \& Carnes, B. A. (1997). Ever since Gompertz.
\emph{Demography}, \emph{34}(1), 1--15.

\leavevmode\hypertarget{ref-ouellette_forme_2016}{}%
Ouellette, N. (2016). La forme de la courbe de mortalité des centenaires
canadiens-français. \emph{Gérontologie et société}, \emph{39}(3),
41--53.

\leavevmode\hypertarget{ref-perks_experiments_1932}{}%
Perks, W. (1932). On some experiments in the graduation of mortality
statistics. \emph{Journal of the Institute of Actuaries}, 12--57.

\leavevmode\hypertarget{ref-pletcher_model_1999}{}%
Pletcher. (1999). Model fitting and hypothesis testing for age-specific
mortality data. \emph{Journal of Evolutionary Biology}, \emph{12}(3),
430--439.

\leavevmode\hypertarget{ref-preston_demography_2000}{}%
Preston, S., Heuveline, P., \& Guillot, M. (2000). \emph{Demography:
Measuring and modeling population processes}.

\leavevmode\hypertarget{ref-promislow_threshold_1999}{}%
Promislow, Tatar, Pletcher, \& Carey. (1999). Below--Threshold
mortality: Implications for studies in evolution, ecology and
demography. \emph{Journal of Evolutionary Biology}, \emph{12}(2),
314--328.

\leavevmode\hypertarget{ref-rcoreteam_language_2014}{}%
R Core Team. (2014). \emph{R: A language and environment for statistical
computing}. Vienna, Austria: R Foundation for Statistical Computing.

\leavevmode\hypertarget{ref-rau_continued_2008}{}%
Rau, R., Soroko, E., Jasilionis, D., \& Vaupel, J. W. (2008). Continued
Reductions in Mortality at Advanced Ages. \emph{Population and
Development Review}, \emph{34}(4), 747--768.

\leavevmode\hypertarget{ref-schloerke_ggally_2014}{}%
Schloerke, B., Crowley, J., Cook, D., Hofmann, H., Wickham, H., Briatte,
F., et al. (2014). \emph{GGally: Extension to ggplot2. R package version
0.5. 0}.

\leavevmode\hypertarget{ref-sheshinski_economic_2007}{}%
Sheshinski, E. (2007). \emph{The Economic Theory of Annuities}.
Princeton University Press.

\leavevmode\hypertarget{ref-slowikowski_ggrepel_2016}{}%
Slowikowski, K., \& Irisson, J. (2016). \emph{Ggrepel: Repulsive Text
and Label Geoms for ggplot2. R package version 0.6. 5}.

\leavevmode\hypertarget{ref-staetsky_diverging_2009}{}%
Staetsky, L. (2009). Diverging trends in female old-age mortality: A
reappraisal. \emph{Demographic Research; Rostock}, \emph{21}, 885--914.

\leavevmode\hypertarget{ref-stein_sage_2008}{}%
Stein, W. (2008). Sage: Open source mathematical software. \emph{7
December 2009}.

\leavevmode\hypertarget{ref-steinsaltz_reevaluating_2005}{}%
Steinsaltz, D. (2005). Re-evaluating a test of the heterogeneity
explanation for mortality plateaus. \emph{Experimental gerontology},
\emph{40}(1-2), 101--113.

\leavevmode\hypertarget{ref-steinsaltz_understanding_2006}{}%
Steinsaltz, D., \& Wachter, K. (2006). Understanding Mortality Rate
Deceleration and Heterogeneity. \emph{Mathematical Population Studies},
\emph{13}(1), 19--37.

\leavevmode\hypertarget{ref-stone_asymptotic_1977}{}%
Stone, M. (1977). An asymptotic equivalence of choice of model by
cross-validation and Akaike's criterion. \emph{Journal of the Royal
Statistical Society. Series B (Methodological)}, 44--47.

\leavevmode\hypertarget{ref-strehler_general_1960}{}%
Strehler, B. L., \& Mildvan, A. S. (1960). General Theory of Mortality
and Aging. \emph{Science}, \emph{132}(3418), 14--21.

\leavevmode\hypertarget{ref-tabeau_forecasting_2001}{}%
Tabeau, E., Van den Berg Jeths, A., \& Heathcote, C. (2001).
\emph{Forecasting mortality in developed countries}. Springer.

\leavevmode\hypertarget{ref-thatcher_force_1998}{}%
Thatcher, A., Kannisto, V., \& Vaupel, J. (1998). \emph{The force of
mortality at ages 80 to 120} (Vol. 22). Odense University Press Odense.

\leavevmode\hypertarget{ref-unitednationspopulationdivision_world_2015}{}%
United Nations Population Division. (2015). \emph{World Population
Prospects: The 2015 Revision, Methodology of the United Nations
Population Estimates and Projections} (No. Working Paper No.
ESA/P/WP.242.).

\leavevmode\hypertarget{ref-vaupel_it_2003}{}%
Vaupel, J. W., Carey, J. R., \& Christensen, K. (2003). It's never too
late. \emph{Science}, \emph{301}(5640), 1679--1681.

\leavevmode\hypertarget{ref-vaupel_deviant_1983}{}%
Vaupel, J., \& Yashin, A. (1983). \emph{The deviant dynamics of death in
heterogeneous populations}. International Institute for Applied Systems
Analysis.

\leavevmode\hypertarget{ref-vaupel_heterogeneity_1985}{}%
Vaupel, J., \& Yashin, A. (1985). Heterogeneity's ruses: Some surprising
effects of selection on population dynamics. \emph{American
statistician}, 176--185.

\leavevmode\hypertarget{ref-vaupel_biodemographic_1998}{}%
Vaupel, J., Carey, J., Christensen, K., Johnson, T., Yashin, A., Holm,
N., et al. (1998). Biodemographic trajectories of longevity.
\emph{Science}, \emph{280}(5365), 855.

\leavevmode\hypertarget{ref-vaupel_impact_1979}{}%
Vaupel, J., Manton, K., \& Stallard, E. (1979). The impact of
heterogeneity in individual frailty on the dynamics of mortality.
\emph{Demography}, \emph{16}(3), 439--454.

\leavevmode\hypertarget{ref-wang_analysis_1998}{}%
Wang, J.-L., Muller, H.-G., \& Capra, W. B. (1998). Analysis of
Oldest-Old Mortality: Lifetables Revisited. \emph{The Annals of
Statistics}, \emph{26}(1), 126--163.

\leavevmode\hypertarget{ref-wasserman_all_2013}{}%
Wasserman, L. (2013). \emph{All of statistics: A concise course in
statistical inference}. Springer Science \& Business Media.

\leavevmode\hypertarget{ref-weibull_statistical_1951}{}%
Weibull, W. (1951). A Statistical Distribution Function of Wide
Applicability. \emph{Journal of applied mechanics}.

\leavevmode\hypertarget{ref-wickham_ggplot2_2009}{}%
Wickham, H. (2009). \emph{Ggplot2: Elegant graphics for data analysis}.
Springer New York.

\leavevmode\hypertarget{ref-wickham_dplyr_2016}{}%
Wickham, H., Francois, R., Henry, L., \& Müller, K. (2016). \emph{Dplyr:
A Grammar of Data Manipulation. R package version 0.5. 0}. R Core
Development Team Vienna.

\leavevmode\hypertarget{ref-wilmoth_are_1995}{}%
Wilmoth, J. R. (1995). Are mortality rates falling at extremely high
ages? An investigation based on a model proposed by Coale and Kisker.
\emph{Population studies}, \emph{49}(2), 281--295.

\leavevmode\hypertarget{ref-wrigley-field_mortality_2014}{}%
Wrigley-Field, E. (2014). Mortality Deceleration and Mortality
Selection: Three Unexpected Implications of a Simple Model.
\emph{Demography}, \emph{51}(1), 51--71.

\leavevmode\hypertarget{ref-yashin_mortality_2000}{}%
Yashin, A. I., Iachine, I. A., \& Begun, A. S. (2000). Mortality
modeling: A review. \emph{Mathematical Population Studies}, \emph{8}(4),
305--332.

\leavevmode\hypertarget{ref-yashin_duality_1994}{}%
Yashin, A., Vaupel, J., \& Iachine, I. (1994). A duality in aging: The
equivalence of mortality models based on radically different concepts.
\emph{Mechanisms of Ageing and Development}, \emph{74}(1-2), 1--14.

\leavevmode\hypertarget{ref-yi_oldest_2003}{}%
Yi, Z., \& Vaupel, J. W. (2003). Oldest old mortality in China.
\emph{Demographic Research}, \emph{8}, 215--244.

\newpage\appendix

\hypertarget{online-appendix}{%
\section*{Online appendix}\label{online-appendix}}
\addcontentsline{toc}{section}{Online appendix}

\hypertarget{sec:ap-data}{%
\section{Data}\label{sec:ap-data}}

The Kannisto-Thatcher database contains data on deaths and exposure
above age 80 for 35 countries, with some of the Scandinavian data going
back to the mid-18th century (MPIDR 2014). However, data quality for
mortality at advanced ages is a serious concern (Jdanov et al. 2008). In
order to assess theories of old-age mortality by fitting hazard models
to cohort data, we narrowed the Kannisto-Thatcher database down,
focusing on cohorts for which very high-quality data mortality is
available. We took several steps to select cohorts of very high quality.
First, we use the analysis of Kannisto-Thatcher data quality reported in
Jdanov et al. (2008) to identify countries for which over half of the
available data years were of the highest quality, and the remaining
years were of the second-highest quality. This set of countries that
consistently produced high-quality data included Belgium, Czech
Republic, Denmark, France, West Germany, Italy, Japan, the Netherlands,
Poland, Scotland, Sweden, and Switzerland. From this subset of countries
that produce high-quality data, we then chose the subset of years that
Jdanov et al. (2008) assessed to have the highest quality.

Because we model deaths at ages 80 to 104, we also require information
about deaths by single year of age in that age range. Many country-years
in the Kannisto-Thatcher database have data reported by single year of
age only up to age 99, 100, or 101; in these country years, an aggregate
category is then used for older deaths. Importantly, the researchers who
curate the Kannisto-Thatcher database assume a Kannisto model in order
to assign deaths in the aggregate category to the individual ages (MPIDR
2014). Using cohorts whose data has been pre-processed using a Kannisto
model would obviously confuse efforts to assess the fit of different
mortality hazards to these data. Thus, we can only use country-years for
which deaths are reported by single year of age up to 105\footnote{For a
  cohort born in year \(c\), the first members will reach their 80th
  birthday in year \(y_{\text{min}} = c+80\) and the last possible age
  at which cohort members could be 104 is \(y_{\text{max}} = c+106\).
  Thus, to be able to model cohort \(c\), I need high quality period
  data from the period \([c+80, c+106)\).}.

Putting these requirements together, Table~\ref{tbl:highqual-countries}
summarizes the cohorts for which high-quality data are available on
deaths by single year of age for the entire time span that the cohort
members were between the ages of 80 and 105. In total, there are 360
country-sex-cohorts of data from 10 different countries.

\hypertarget{tbl:highqual-countries}{}
\begin{longtable}[]{@{}lllll@{}}
\caption{\label{tbl:highqual-countries}Summary of the availability of
high-quality data from the Kannisto-Thatcher database. Usable ages means
that deaths are reported by single year of age up to at least age 105.
The final dataset has data for 360 country-sex-cohorts.}\tabularnewline
\toprule
\begin{minipage}[b]{0.12\columnwidth}\raggedright
Country\strut
\end{minipage} & \begin{minipage}[b]{0.32\columnwidth}\raggedright
High quality years\strut
\end{minipage} & \begin{minipage}[b]{0.13\columnwidth}\raggedright
Usable ages\strut
\end{minipage} & \begin{minipage}[b]{0.16\columnwidth}\raggedright
Usable periods\strut
\end{minipage} & \begin{minipage}[b]{0.13\columnwidth}\raggedright
Cohorts included\strut
\end{minipage}\tabularnewline
\midrule
\endfirsthead
\toprule
\begin{minipage}[b]{0.12\columnwidth}\raggedright
Country\strut
\end{minipage} & \begin{minipage}[b]{0.32\columnwidth}\raggedright
High quality years\strut
\end{minipage} & \begin{minipage}[b]{0.13\columnwidth}\raggedright
Usable ages\strut
\end{minipage} & \begin{minipage}[b]{0.16\columnwidth}\raggedright
Usable periods\strut
\end{minipage} & \begin{minipage}[b]{0.13\columnwidth}\raggedright
Cohorts included\strut
\end{minipage}\tabularnewline
\midrule
\endhead
\begin{minipage}[t]{0.12\columnwidth}\raggedright
Belgium\strut
\end{minipage} & \begin{minipage}[t]{0.32\columnwidth}\raggedright
1951-2000\strut
\end{minipage} & \begin{minipage}[t]{0.13\columnwidth}\raggedright
1973+\strut
\end{minipage} & \begin{minipage}[t]{0.16\columnwidth}\raggedright
1973-2000\strut
\end{minipage} & \begin{minipage}[t]{0.13\columnwidth}\raggedright
1893-1895\strut
\end{minipage}\tabularnewline
\begin{minipage}[t]{0.12\columnwidth}\raggedright
Czech Republic\strut
\end{minipage} & \begin{minipage}[t]{0.32\columnwidth}\raggedright
1951-1980\strut
\end{minipage} & \begin{minipage}[t]{0.13\columnwidth}\raggedright
none\strut
\end{minipage} & \begin{minipage}[t]{0.16\columnwidth}\raggedright
none\strut
\end{minipage} & \begin{minipage}[t]{0.13\columnwidth}\raggedright
none\strut
\end{minipage}\tabularnewline
\begin{minipage}[t]{0.12\columnwidth}\raggedright
Denmark\strut
\end{minipage} & \begin{minipage}[t]{0.32\columnwidth}\raggedright
1931-1950; 1961-2000\strut
\end{minipage} & \begin{minipage}[t]{0.13\columnwidth}\raggedright
1943+\strut
\end{minipage} & \begin{minipage}[t]{0.16\columnwidth}\raggedright
1943-1950; 1961-2000\strut
\end{minipage} & \begin{minipage}[t]{0.13\columnwidth}\raggedright
1881-1895\strut
\end{minipage}\tabularnewline
\begin{minipage}[t]{0.12\columnwidth}\raggedright
France\strut
\end{minipage} & \begin{minipage}[t]{0.32\columnwidth}\raggedright
1951-2000\strut
\end{minipage} & \begin{minipage}[t]{0.13\columnwidth}\raggedright
1946-1997;2002+\strut
\end{minipage} & \begin{minipage}[t]{0.16\columnwidth}\raggedright
1946-1997\strut
\end{minipage} & \begin{minipage}[t]{0.13\columnwidth}\raggedright
1866-1892\strut
\end{minipage}\tabularnewline
\begin{minipage}[t]{0.12\columnwidth}\raggedright
W. Germany\strut
\end{minipage} & \begin{minipage}[t]{0.32\columnwidth}\raggedright
1971-2000\strut
\end{minipage} & \begin{minipage}[t]{0.13\columnwidth}\raggedright
1964+\strut
\end{minipage} & \begin{minipage}[t]{0.16\columnwidth}\raggedright
1971-2000\strut
\end{minipage} & \begin{minipage}[t]{0.13\columnwidth}\raggedright
1891-1895\strut
\end{minipage}\tabularnewline
\begin{minipage}[t]{0.12\columnwidth}\raggedright
Italy\strut
\end{minipage} & \begin{minipage}[t]{0.32\columnwidth}\raggedright
1961-2000\strut
\end{minipage} & \begin{minipage}[t]{0.13\columnwidth}\raggedright
all\strut
\end{minipage} & \begin{minipage}[t]{0.16\columnwidth}\raggedright
1961-2000\strut
\end{minipage} & \begin{minipage}[t]{0.13\columnwidth}\raggedright
1881-1895\strut
\end{minipage}\tabularnewline
\begin{minipage}[t]{0.12\columnwidth}\raggedright
Japan\strut
\end{minipage} & \begin{minipage}[t]{0.32\columnwidth}\raggedright
1971-2000\strut
\end{minipage} & \begin{minipage}[t]{0.13\columnwidth}\raggedright
all\strut
\end{minipage} & \begin{minipage}[t]{0.16\columnwidth}\raggedright
1971-2000\strut
\end{minipage} & \begin{minipage}[t]{0.13\columnwidth}\raggedright
1891-1895\strut
\end{minipage}\tabularnewline
\begin{minipage}[t]{0.12\columnwidth}\raggedright
Netherlands\strut
\end{minipage} & \begin{minipage}[t]{0.32\columnwidth}\raggedright
1951-2000\strut
\end{minipage} & \begin{minipage}[t]{0.13\columnwidth}\raggedright
all\strut
\end{minipage} & \begin{minipage}[t]{0.16\columnwidth}\raggedright
1951-2000\strut
\end{minipage} & \begin{minipage}[t]{0.13\columnwidth}\raggedright
1871-1895\strut
\end{minipage}\tabularnewline
\begin{minipage}[t]{0.12\columnwidth}\raggedright
Poland\strut
\end{minipage} & \begin{minipage}[t]{0.32\columnwidth}\raggedright
1971-2000\strut
\end{minipage} & \begin{minipage}[t]{0.13\columnwidth}\raggedright
2002+\strut
\end{minipage} & \begin{minipage}[t]{0.16\columnwidth}\raggedright
none\strut
\end{minipage} & \begin{minipage}[t]{0.13\columnwidth}\raggedright
none\strut
\end{minipage}\tabularnewline
\begin{minipage}[t]{0.12\columnwidth}\raggedright
Scotland\strut
\end{minipage} & \begin{minipage}[t]{0.32\columnwidth}\raggedright
1951-1960; 1971-1990\strut
\end{minipage} & \begin{minipage}[t]{0.13\columnwidth}\raggedright
1963-2003\strut
\end{minipage} & \begin{minipage}[t]{0.16\columnwidth}\raggedright
1971-1990\strut
\end{minipage} & \begin{minipage}[t]{0.13\columnwidth}\raggedright
1891-1895\strut
\end{minipage}\tabularnewline
\begin{minipage}[t]{0.12\columnwidth}\raggedright
Sweden\strut
\end{minipage} & \begin{minipage}[t]{0.32\columnwidth}\raggedright
1901-2000\strut
\end{minipage} & \begin{minipage}[t]{0.13\columnwidth}\raggedright
all\strut
\end{minipage} & \begin{minipage}[t]{0.16\columnwidth}\raggedright
1901-2000\strut
\end{minipage} & \begin{minipage}[t]{0.13\columnwidth}\raggedright
1821-1895\strut
\end{minipage}\tabularnewline
\begin{minipage}[t]{0.12\columnwidth}\raggedright
Switzerland\strut
\end{minipage} & \begin{minipage}[t]{0.32\columnwidth}\raggedright
1911-1920; 1931-1940; 1951-1960; 1971-2000\strut
\end{minipage} & \begin{minipage}[t]{0.13\columnwidth}\raggedright
1950+\strut
\end{minipage} & \begin{minipage}[t]{0.16\columnwidth}\raggedright
1951-1960; 1971-2000\strut
\end{minipage} & \begin{minipage}[t]{0.13\columnwidth}\raggedright
1891-1895\strut
\end{minipage}\tabularnewline
\bottomrule
\end{longtable}

\hypertarget{methods}{%
\section{Methods}\label{methods}}

This appendix provides greater detail about the models and the methods
used to fit each model to cohort death data. Appendix
\ref{sec:ap-heterogeneity} provides a more detailed explanation of the
theory behind mortality hazards. Next, Appendix \ref{sec:ap-model}
reviews the model that was introduced in the main text. Finally,
Appendix \ref{sec:ap-estimation} provides some technical details related
to fitting the hazard models by maximum likelihood.

\hypertarget{sec:ap-heterogeneity}{%
\subsection{Hazards and heterogeneity}\label{sec:ap-heterogeneity}}

Models of mortality at old ages can be expressed using hazard functions,
which are mathematical descriptions of the risk of death by age.
Researchers distinguish between the individual hazard function and the
population hazard function. For a particular person, \(i\), the
\emph{individual hazard function} is defined as

\[
\mu_i(z) = -\frac{d \log S_i(z)}{d z} = - \frac{1}{S_i(z)} \frac{d S_i(z)}{d z}.
\]

\noindent where \(S_i(z)\) is the \emph{individual survival function},
\emph{i.e.} the probability that person \(i\) is alive at exact age
\(z\).

The individual hazard function is a useful theoretical concept, but it
can never be empirically observed.\footnote{In order to obtain empirical
  information about the shape of a particular individual's mortality
  hazard function, it would be necessary to repeat the individual's life
  many times, recording the age at which the individual died each time.
  Of course, this is impossible.} Instead, researchers typically study
deaths among groups of people. For a group of people, the
\emph{population hazard function} is defined as:

\begin{equation}
    \mu(z) = -\frac{d \log S(z)}{d z} = - \frac{1}{S(z)} \frac{d S(z)}{d z}. 
\label{eq:ap-popnhaz}\end{equation}

\noindent where \(S(z)\) is the \emph{population survival function},
\emph{i.e.} the proportion of group members that survives to exact age
\(z\).

Mortality researchers are typically interested in individual-level
hazards, but it is only possible to empirically estimate
population-level hazards. What is the relationship between the
individual and population hazard functions? If all of the people in a
group face identical mortality hazards, then the population hazard
function is the same as the group members' individual hazard function;
intuitively, when all group members face identical mortality hazards,
observing the group members' ages at death is equivalent to observing
several lifetimes for one individual. On the other hand, if the people
in a group face different individual hazards, then the population hazard
function is not necessarily the same as any individual hazard; in fact,
research on heterogeneity in mortality has shown that the population
hazard can present a misleading picture of the underlying individuals'
hazard functions (Vaupel and Yashin 1983, 1985; Vaupel et al. 1979;
Wrigley-Field 2014; A. I. Yashin et al. 2000; A. Yashin et al. 1994).
This stream of research has led to many different theories about how
individual hazards might aggregate up into the population hazard.
However, research on mortality heterogeneity has also led to the
understanding that different combinations of individual hazard functions
can lead to the same population hazard function (A. Yashin et al. 1994).
Thus, studying the predictions made by population hazard functions about
cohort deaths by age enables researchers to distinguish between some --
but not all -- theories about mortality; it is technically possible for
different models to predict identical numbers of cohort deaths by age.
When two different models predict exactly the same number of deaths by
age, there is no way to distinguish between them without additional
information. Thus, our design enables us to understand which
\emph{population hazards} best fit available data, but it does not
enable us to distinguish between different theories that predict
identical population hazards (Figure~\ref{fig:theory-and-agg}).

Given a model for the population hazard, there is a mathematical
relationship that links the hazard function \(\mu(z)\) and the
probability of dying between ages \(z\) and \(z+k\), conditional on
surviving to age \(z\):

\begin{equation}
\pi(z,z+k) = 1 - \exp\left( - \int_{z}^{z+k} \mu(x) dx\right).
\label{eq:ap-deathprob}\end{equation}

\noindent Equation~\ref{eq:ap-deathprob} shows that a population hazard
function can be converted into expected numbers of cohort deaths by age.
A theory that has been expressed as a hazard function can thus be tested
by comparing predicted numbers of deaths by age to empirically observed
numbers of deaths by age.

\hypertarget{sec:ap-model}{%
\subsection{Model}\label{sec:ap-model}}

In a population that faces identical hazards cohort deaths between ages
\(z\) and \(z+1\) are distributed binomially; that is, if \(N_z\) people
from a cohort survive to exact age \(z\), and all of the members of the
cohort face the same hazard \(\mu(z)\), then

\begin{equation}
D_z \sim \text{Binomial}( N_z, \pi(z, z+1)),
\label{eq:ap-binomialmodel}\end{equation}

\noindent where \(D_z\) is the number of deaths between ages \(z\) and
\(z+1\), and \(\pi(z,z+1)\) is the probability of dying between ages
\(z\) and \(z+1\). \(\pi(z,z+1)\) is derived from the hazard function
according to:

\begin{equation}
\pi(z,z+k) = 1 - \exp\left( - \int_{z}^{z+k} \mu(x) dx\right).
\label{eq:ap-deathprob}\end{equation}

\noindent The likelihood for an observed sequence of deaths
\(\mathbf{D} = D_1, D_2, \dots\) and survivors to each age,
\(\mathbf{N} = N_1, N_2, \dots\) is then

\[Pr(\mathbf{D} | \mathbf{z}, \theta,\mathbf{N}) = \prod_z {{N_z}\choose{D_z}} \pi(z,\theta)^{D_z} (1-\pi(z,\theta))^{(N_z - D_z)}.\]

Taking logs produces

\begin{equation}
  ll(\mathbf{D}| \mathbf{z}, \theta, \mathbf{N})) = K + \sum_z\left[
    D(z)\log(\pi(z, \theta)) + (N(z)-D(z))\log(1-\pi(z,
    \theta))\right],
\label{eq:ap-ll}\end{equation}

\noindent where \(K\) is a constant that does not vary with \(\theta\).
Finally, substituting in the relationship between the hazard function
and the probability of death, we arrive at

\begin{equation}
\begin{aligned}
  ll(\mathbf{D}| \mathbf{z},\mu, \theta) &= K +\\
   &~~~\sum_z\left[ D(z)\log\left( 1 - \exp\left( - \int_{z}^{z+k} \mu(x, \theta) dx\right)\right) + \right. \\
   &~~~\left.~~~~~(N(z)-D(z)) \exp\left( - \int_{z}^{z+k} \mu(x, \theta) dx\right) \right],
\end{aligned}
\label{eq:ap-ll-full}\end{equation}

In order to fit each model to a cohort dataset,
Equation~\ref{eq:ap-ll-full} is maximized as a function of \(\theta\).

\hypertarget{sec:ap-estimation}{%
\subsection{Estimation}\label{sec:ap-estimation}}

In order to fit a model to a specific cohort dataset, we maximize the
likelihood (Equation~\ref{eq:ap-ll}) numerically using the
Broyden-Fletcher-Goldfarb-Shanno (BFGS) algorithm (Nocedal and Wright
1999). The BFGS algorithm is a numerical optimization routine that has
been shown to work well for different settings and that is currently
considered the most effective of the so-called quasi-Newton optimization
routines (Nocedal and Wright 1999, pg. 139). Since the size of the
problem we study here is not very large, we have no need to adopt
limited memory BFGS, stochastic gradient descent, or other optimization
routines that have become popular for problems with very large numbers
of free parameters. BFGS has been implemented and widely used as part of
the \texttt{optim} routine in the \texttt{R} statistical software
package (R Core Team 2014). The exact parameterization used for each
model is described in Appendix \ref{sec:hazards}.

In maximizing log-likelihoods, it is common to omit constant factors
that do not vary with model parameters. In general, in order to
calculate AIC and BIC values that compare different probability models,
it is important to retain these constant factors. In this study, the
probability model is the same for all of the models---\emph{i.e.}, given
conditional probabilities of death, the deaths are distributed
binomially. Thus, in calculating \(\Delta\text{AIC}\) and
\(\Delta\text{BIC}\), these constants cancel out.

We took several steps to ensure that the BFGS optimization routine
behaved well. First, for each model, we derived closed-form
expressions\footnote{The hazard to probability integral for the
  log-quadratic model involves the error function (Abramowitz and Stegun
  1964; Stein 2008); thus, this result is technically not in closed
  form. However, the error function has been widely studied and highly
  accurate approximations are available.} for (1) the exponentiated
integral that converts hazards to probabilities
(Equation~\ref{eq:deathprob})\footnote{Many previous studies have used
  approximations to convert the hazard into conditional probabilities of
  death, particularly in the context of regression approaches to fitting
  mortality models. Wang et al. (1998) points out that these
  approximations can be problematic, particularly at oldest ages. Thus,
  an additional advantage to deriving analytical expressions for the
  conditional probability of death is that it avoids the need to make
  such approximations.}; (2) the partial derivative of the likelihood
with respect to each parameter value (\emph{i.e.}, partial derivatives
of Equation~\ref{eq:ap-ll} with respect to each entry of the vector
\(\theta\)). Several of the mathematical expressions that result are
complex and unwieldy; thus, to minimize the risk of errors in
translating the mathematical expressions into computer code, we used the
Sage symbolic computing package to programmatically generate C++ code
needed to calculate each analytical expression (Stein 2008). All of the
generated code can be seen in the open-source \texttt{mortfit} package,
and the analytic expression for converting hazards into conditional
probabilities of death are shown in Appendix \ref{sec:hazards}.

Second, in order to guard against the possibility that the BFGS
algorithm converges prematurely or to a local optimum, we maximize the
likelihood many different times (11 in total). The first time we
maximize the likelihood, we use a heuristic to crudely estimate initial
parameter values. The heuristic used is different for each model and a
detailed explanation of each heuristic is described along with the
individual hazard functions in Appendix \ref{sec:hazards}. For each of
the remaining 10 times the BFGS routine is used to maximize the
likelihood, we choose random starting values for the parameters. The
support of the random starting values was determined from initial runs
of the model fits; these initial runs and experiments also guide the
parameter scaling. After estimating \(\theta\) for eleven different sets
of starting values, we pick the result that has the highest value of the
log-likelihood.

Finally, as we developed the routines used to fit the models, we also
added the capability to examine likelihood profile plots and to fit
models using grid searches to the \texttt{mortfit} package. We found
likelihood profile plots to be helpful in initial development of the
optimization strategy, but we did not find that grid search added value
above and beyond using many random starts. Further, grid search can
demand intensive computing resources; thus, we did not use grid search
for the final results.

\hypertarget{sec:ap-results}{%
\section{Additional empirical results}\label{sec:ap-results}}

This Appendix provides additional empirical results based on
alternative, principled approaches to assessing model fit. We review two
additional approaches to model selection: Section~\ref{sec:ap-bic}
discusses the BIC; and Section~\ref{sec:ap-cv} discusses \(K\)-fold
cross validation. The goal of this study is not to argue that one of
these methods is definitively more appropriate than the others; all
three of these approaches have been the subject of theoretical and
empirical motivation, and each has merits and limitations.

\hypertarget{sec:ap-bic}{%
\subsection{Bayesian Information Criterion (BIC)}\label{sec:ap-bic}}

The BIC can be written as

\begin{equation}
  \text{BIC} = -2  \mathfrak{L} + \log(n) k,
\label{eq:bic}\end{equation}

\noindent where \(\mathfrak{L}\) is the value of the maximized
log-likelihood, \(k\) is the number of parameters (\emph{i.e}, the
dimension of \(\theta\)), and \(n\) is the size of the cohort. As was
the case with the AIC, the BIC looks relatively simple, but it is
justified by rigorous theory. Intuitively, the BIC is similar to the
AIC, except that the BIC multiplies the number of parameters in the
model by \(\log(n)\) instead of 2. Since \(\log(n) > 2\) as long as
\(n\) is at least 8, this means that the BIC penalizes complex models
more heavily than the AIC.

The theory that motivates the BIC usually posits a Bayesian setup with
weak priors on the model parameter values and uniform priors over the
candidate models. If the true, data-generating model is in the candidate
set, one can show that the BIC is then consistent in the sense that the
BIC will tend to select the true data-generating model with probability
1 as the sample size \(n\) goes to infinity. However, it seems
unrealistic to think that the true data-generating model will ever
really be in the candidate set; the more relevant question seems to be:
which of the candidate models makes the most accurate empirical
predictions? Viewed from that perspective, the BIC is a model selection
criterion that more heavily penalizes complex models than the AIC;
indeed, some evidence suggest that the BIC can favor overly simple
models (Burnham and Anderson 2004; Claeskens and Hjort 2008, Ch. 4;
Efron and Hastie 2016, Ch. 13).

Like the AIC, absolute BIC values are not comparable across datasets,
but relative values are. Thus, following Burnham and Anderson (2004), we
use the quantity \(\Delta\text{BIC}\), defined in the same way as
\(\Delta\text{AIC}\), to compare model fits across all cohorts in our
sample.

More detailed discussions of the BIC, including comparisons with the
AIC, can be found in Claeskens and Hjort (2008); Burnham and Anderson
(2003); Burnham and Anderson (2004); Hoeting et al. (1999); Friedman et
al. (2009); and Kass and Raftery (1995).

\hypertarget{results-using-the-bic}{%
\subsubsection*{Results using the BIC}\label{results-using-the-bic}}
\addcontentsline{toc}{subsubsection}{Results using the BIC}

In this section, we present summaries of model fits across the 360
cohorts as evaluated by the BIC. In general, the BIC can be expected to
prefer models with fewer parameters than those selected by the AIC; in
the results here, this preference for less complex models is most
evident in the comparatively better performance of the two-parameter
Kannisto and Gompertz models.

Appendix Figure~\ref{fig:deltabicboxplot} shows boxplots that summarize
the \(\Delta\text{BIC}\) values by model across all 360 cohorts; this
figure is analogous to Figure~\ref{fig:deltaaicboxplot}. According the
\(\Delta\text{BIC}\), the two-parameter Kannisto model performs the
best.

\hypertarget{fig:deltabicboxplot}{%
\begin{figure}
\centering
\includegraphics{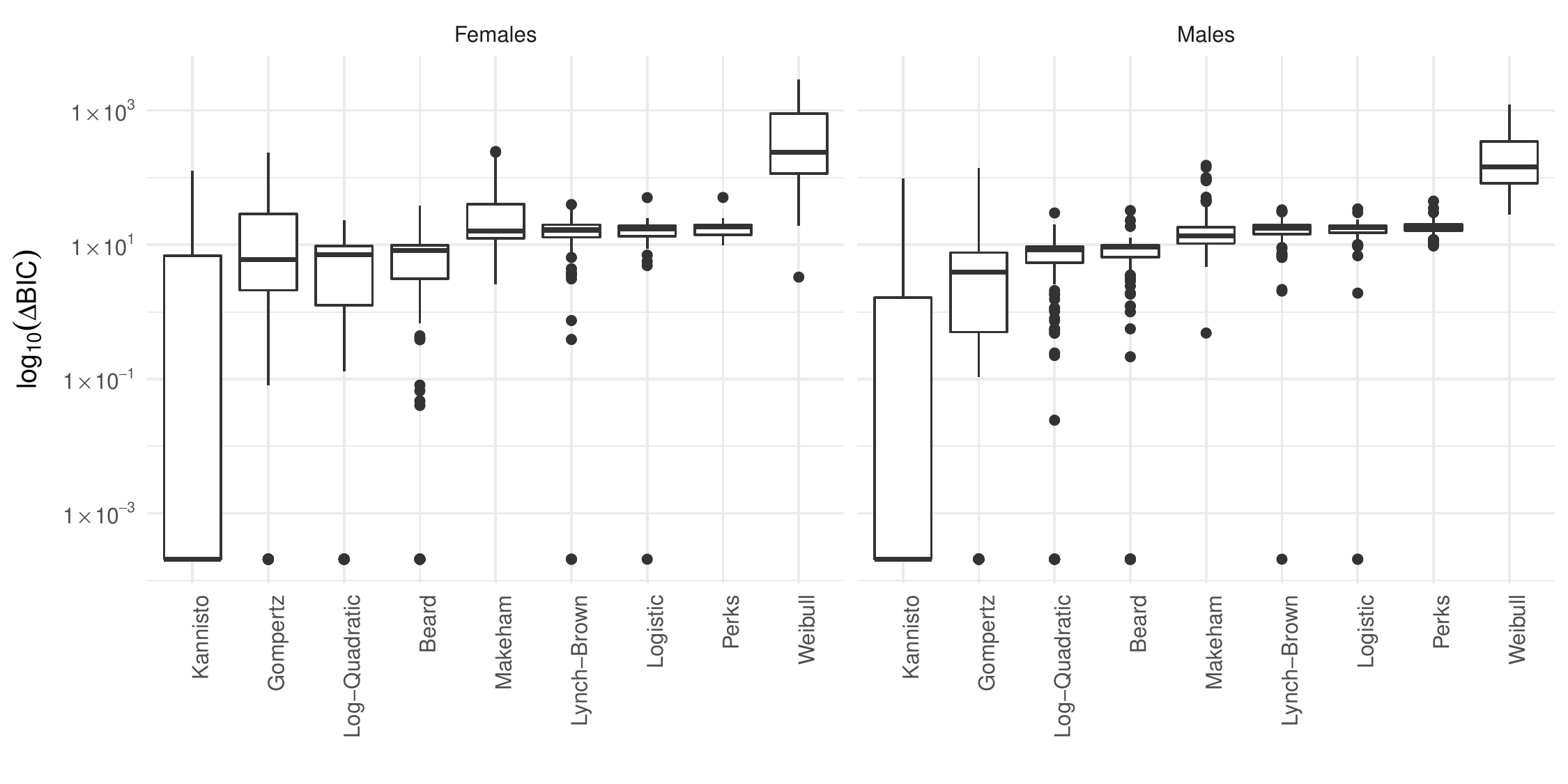}
\caption{Boxplots summarizing the distributions of \(\Delta\text{BIC}\)
across all of the cohorts in the sample (note that the y axis is on a
log scale). The horizontal line shows the median, the edges of the box
show the interquartile range, and the whiskers extend to the largest and
smallest values within 1.5 times the interquartile range; more extreme
points are plotted separately. The Kannisto model appears to perform the
best.}\label{fig:deltabicboxplot}
\end{figure}
}

Appendix Figure~\ref{fig:deltabicgoodbad} shows \(\Delta\text{BIC}\)
results analogous to the \(\Delta\text{AIC}\) results presented in
Figure~\ref{fig:deltaaicgoodbad}. Appendix
Figure~\ref{fig:deltabicgoodbad} compares the fraction of cohorts for
which each model enjoys substantial support from the data
(\(\Delta\text{BIC} \leq 2\)) to the fraction of cohorts for which each
model has almost no support from the data (\(\Delta\text{BIC} > 10\)).
The figure dramatically separates models according to the number of
parameters they have. The Kannisto model is the best performer, with
substantial support from the data (\(\Delta\text{BIC} \leq 2\)) in about
three quarters of the cohorts, and essentially no support from the data
(\(\Delta\text{BIC} > 10\)) in fewer than one quarter of the cohorts.
The Gompertz model (2 parameters) and the Log-Quadratic and Beard models
(3 parameters) are in the lower-left corner of the plot: these models
infrequently have substantial support from the data
(\(\Delta\text{BIC} \leq 2\)), but they also infrequently have
essentially no support from the data (\(\Delta\text{BIC} > 10\)). All of
the remaining models are in the top-left corner of the plot, meaning
that \(\Delta\text{BIC}\) suggests that they have little empirical
support.

\hypertarget{fig:deltabicgoodbad}{%
\begin{figure}
\centering
\includegraphics{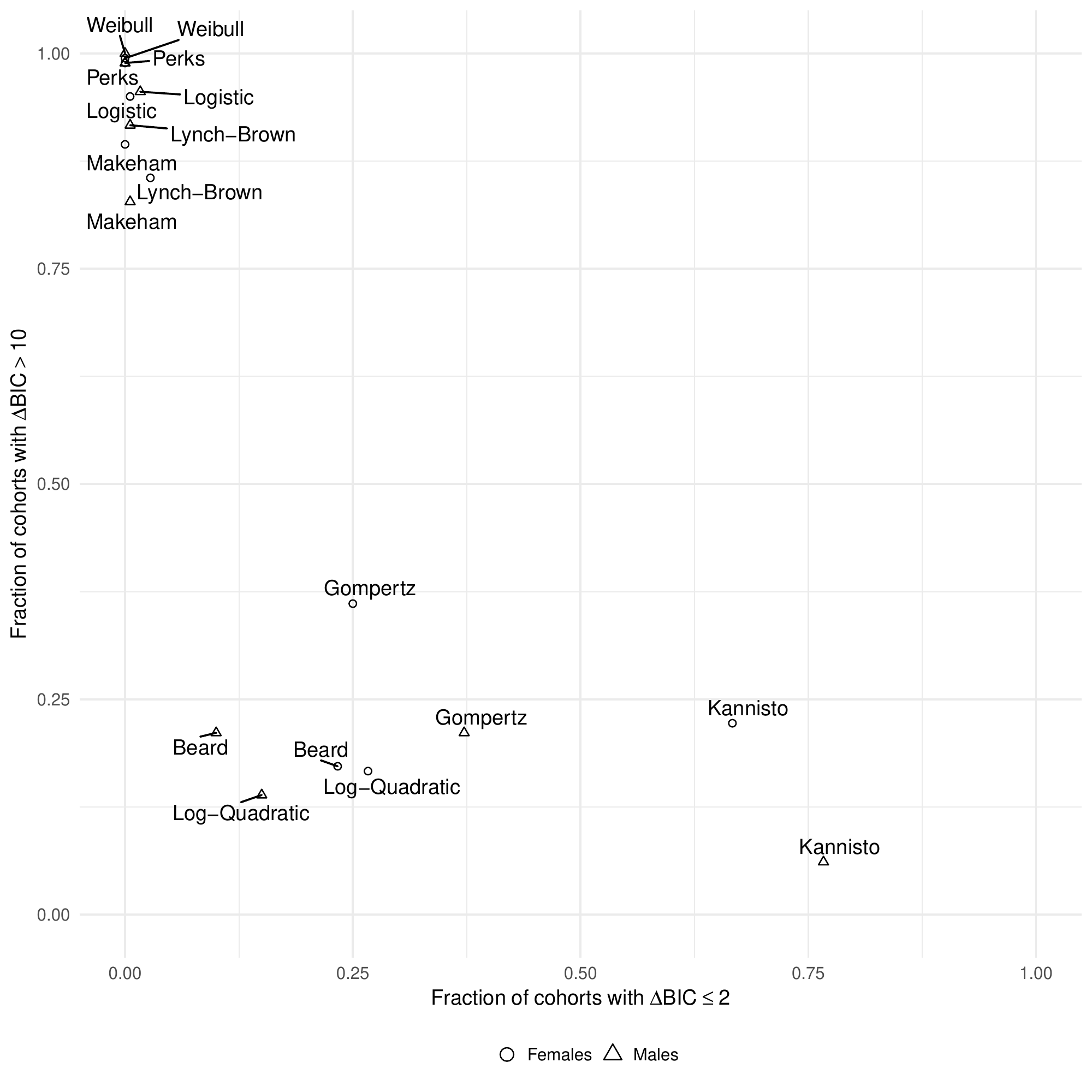}
\caption{Fraction of cohorts for which the fit of each model had
substantial support (\(\Delta\text{BIC} <= 2\), x axis) compared to the
fraction of cohorts for which the fit of each model had essentially no
support (\(\Delta\text{BIC} > 10\), y axis). Models in the lower-right
corner consistently fit the cohorts in the sample well; models in the
lower-left corner are neither very good nor very bad; models in the
upper-left corner consistently fit the cohorts in the sample
poorly.}\label{fig:deltabicgoodbad}
\end{figure}
}

Appendix Figure~\ref{fig:deltabicgoodbadcountry} shows
\(\Delta\text{BIC}\) results analogous to the \(\Delta\text{AIC}\)
results presented in Figure~\ref{fig:deltaaicgoodbadcountry}. Appendix
Figure~\ref{fig:deltabicgoodbadcountry} breaks the results in Appendix
Figure~\ref{fig:deltabicgoodbad} down by country for the five countries
in the sample that have more than 10 cohorts worth of data. The figure
shows that for Denmark, the Netherlands, and Sweden, the Kannisto model
performs well. For Italy, and for French females, the Log-Quadratic
model is preferred. More generally, consistent with the findings from
Appendix Figure~\ref{fig:deltabicgoodbad}, results using
\(\Delta\text{BIC}\) heavily penalize more complex models: models with
four parameters (Logistic and Lynch-Brown) never perform well.

\hypertarget{fig:deltabicgoodbadcountry}{%
\begin{figure}
\centering
\includegraphics[width=\textwidth,height=1.1\textwidth]{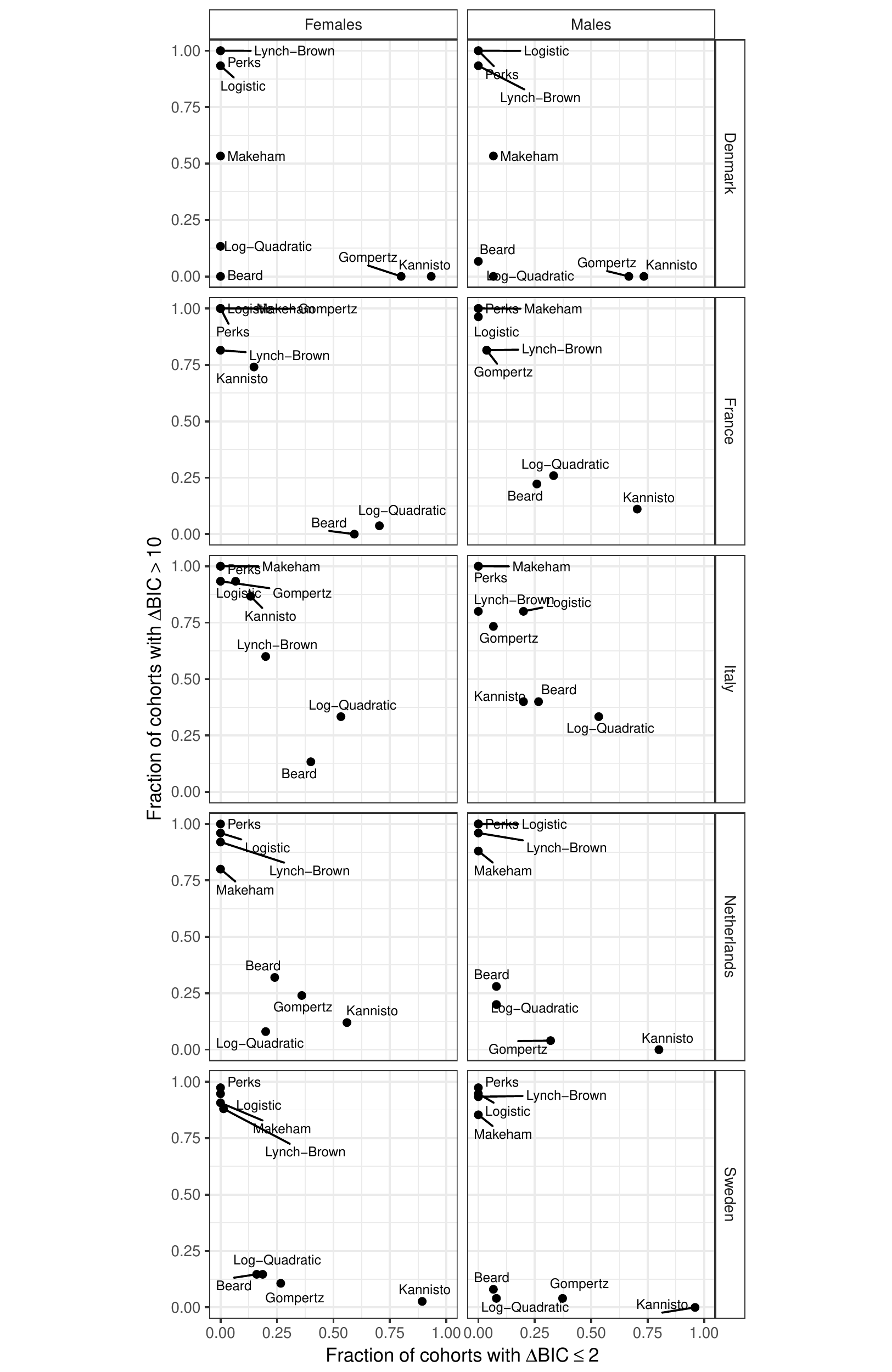}
\caption{Fraction of cohorts for which the fit of each model had
substantial support (\(\Delta\text{BIC} <= 2\), x axis) compared to the
fraction of cohorts for which the fit of each model had essentially no
support (\(\Delta\text{BIC} > 10\), y axis) for each sex and for each
country with at least ten cohorts of data. Models in the lower-right
corner consistently fit the cohorts in the sample well; models in the
lower-left corner are neither very good nor very bad; models in the
upper-left corner consistently fit the cohorts in the sample
poorly.}\label{fig:deltabicgoodbadcountry}
\end{figure}
}

\hypertarget{sec:ap-cv}{%
\subsection{\texorpdfstring{\(K\)-fold cross
validation}{K-fold cross validation}}\label{sec:ap-cv}}

\(K\)-fold cross validation is based on the idea that a dataset can be
randomly split into \(K\) different parts, called \emph{folds}. The
model being studied can then be fit to \emph{K-1} of the folds (the
\emph{training set}), and the fitted model can be used to predict the
outcome---\emph{i.e.}, the number of deaths by age---for the remaining
fold (the \emph{test set}). Intuitively, this process overcomes the
problem of overfitting because the parameters are estimated from one
part of the data, and then the accuracy is assessed on a different part
of the data. Thus, if stochastic noise from the training set had a large
influence on parameter estimates, that will lead to error when the
fitted model is used to predict deaths in the test set. In \(K\)-fold
cross validation, this process can be repeated \(K\) times, each time
picking a different fold to serve as the hold-out sample. Overall
measures of prediction error can then be averaged over the \(K\)
different cross-validation error estimates. Friedman et al. (2009),
Wasserman (2013), and Efron and Hastie (2016) have introductions to
\(K\)-fold cross validation.

Cross validation has been the subject of somewhat less theoretical
analysis than AIC and BIC, but it is appealing because it can be easily
used for a wide range of different models, including nonparametric
models and prediction algorithms that are not even based on probability
distributions. Cross validation has become increasingly popular as
computation has become relatively cheap and because it has been very
influential in machine learning. To the extent that cross validation has
been theoretically studied, its motivation is similar to the AIC; for
example, leave-one-out cross validation has been shown to be
asymptotically equivalent to the AIC (Burnham and Anderson 2003;
Claeskens and Hjort 2008; Stone 1977). More generally, the AIC and cross
validation both aim to reflect a model's predictive ability. Cross
validation is extremely flexible, while the AIC is more efficient in the
sense that it provides less noisy measures of predictive ability (Efron
and Hastie 2016, Ch. 12).

For cohort death data, the number of observations is given by the size
of the cohort at the start of the youngest age group. The
\texttt{mortfit} package has functions that can be used to split cohort
death datasets into \(K\) folds, and the model estimation routines
described in Appendix \ref{sec:ap-estimation} can be repeated for each
fold. For example, within each fold, the model is fit with 11 different
starting values, one based on heuristics and ten picked randomly. The
results are used to estimate prediction error. The results here use
\(K=5\) folds.

\hypertarget{results-using-k-fold-cross-validation}{%
\subsection*{Results using K-fold
cross-validation}\label{results-using-k-fold-cross-validation}}
\addcontentsline{toc}{subsection}{Results using K-fold cross-validation}

In this section, we present summaries of model fits across the 360
cohorts as evaluated by 5-fold cross validation. In general, theoretical
results show that cross validation can be expected to prefer similar
models to the AIC, though the AIC is more efficient (Appendix
\ref{sec:ap-cv}).

For \(K\)-fold cross validation, there is no direct analog to
\(\Delta\text{AIC}\). Thus, we present results based on the rank of each
model within each cohort. Since the Weibull model consistently performs
very poorly, we omit it from these summaries. Thus, within each cohort,
model ranks range from 1 (best fit) to 8 (worst fit).

Appendix Figure~\ref{fig:cvrank} shows the distribution of ranks across
the 360 cohort fits for each model, and for each sex. According to the
median rank, the top-performing models are the Lynch-Brown, Logistic,
Beard, and Log-Quadratic; the bottom performers are the Gompertz and
Makeham models. As would be expected from statistical theory, the
results in Appendix Figure~\ref{fig:cvrank} suggest that \(K\)-fold
cross validation penalizes complex models less heavily than the BIC, and
more similarly to the AIC.

\hypertarget{fig:cvrank}{%
\begin{figure}
\centering
\includegraphics{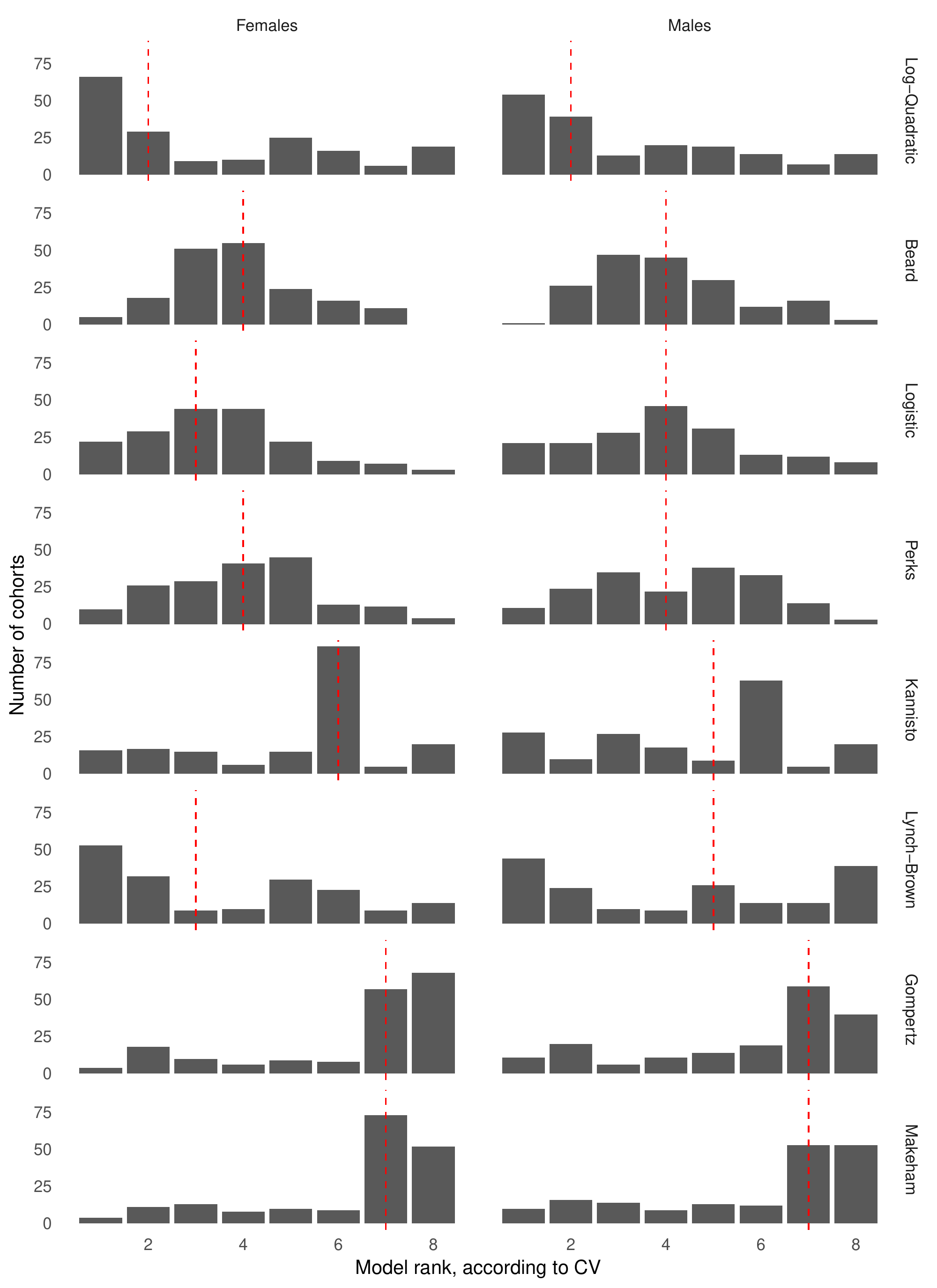}
\caption{Distribution of model rank by 5-fold cross-validation, with 1
being the best prediction and 8 being the worst. Male cohorts are in the
left column and female cohorts are in the right column. The dashed
vertical lines show the median rank for each model.}\label{fig:cvrank}
\end{figure}
}

Appendix Figure~\ref{fig:cvgoodbad} compares the fraction for cohorts
for which each model is in the top 2 (x axis) to the fraction of cohorts
for which each model is in the bottom two (y axis). The figure shows
that, according to 5-fold cross validation, the Makeham and Gompertz
models consistently perform more poorly than all of the others. The
remaining models are remarkably similar to one another in terms of their
performance; the point estimates in Appendix Figure~\ref{fig:cvgoodbad}
suggest that the Lynch-Brown and Log-Quadratic models attain a good
trade-off between consistently fitting relatively well and rarely
fitting relatively poorly.

\hypertarget{fig:cvgoodbad}{%
\begin{figure}
\centering
\includegraphics{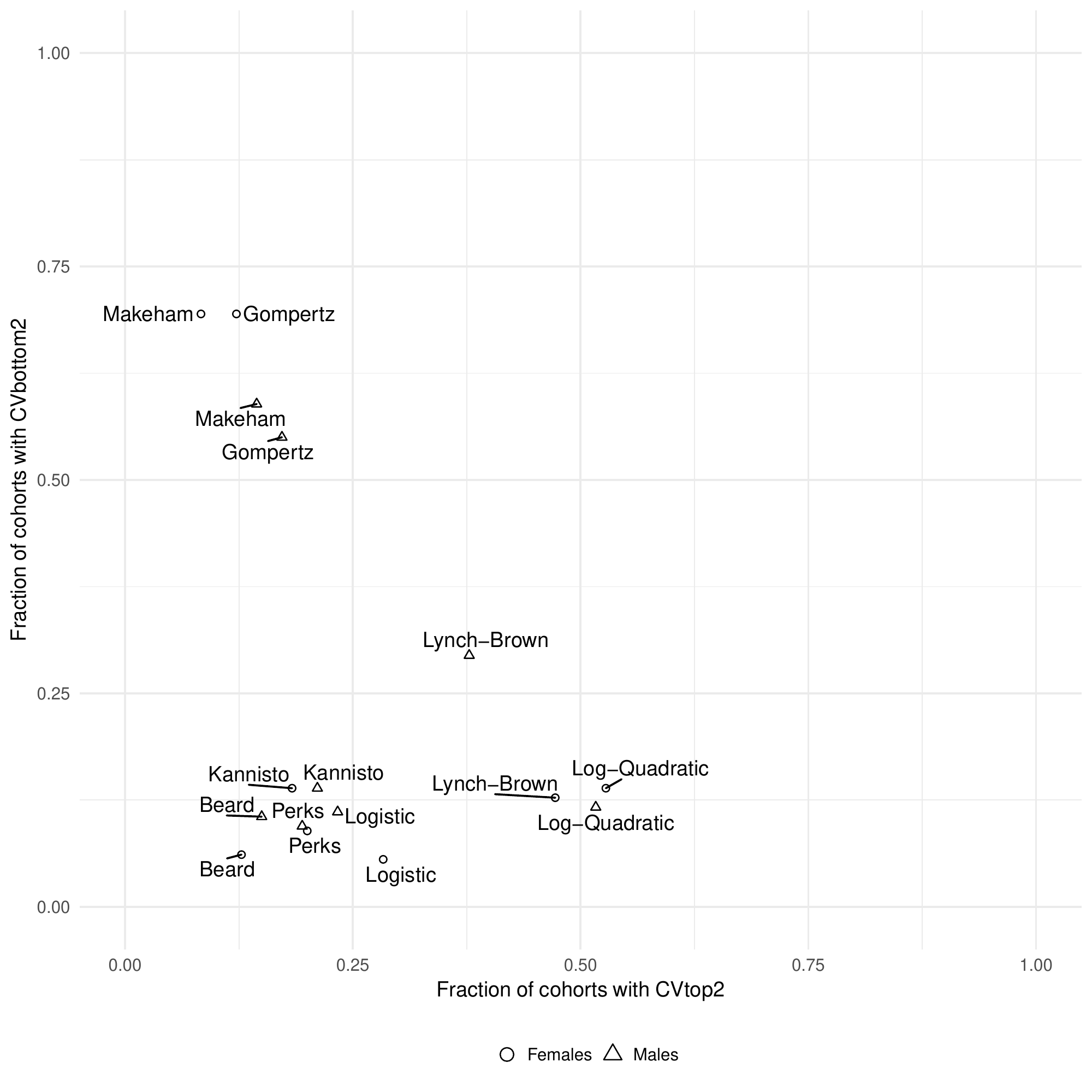}
\caption{Fraction of cohorts for which the cross-validation error of
each model was very good (best or second-best, x axis) and very poor
(worst or second-worst, y axis). Models in the lower-right corner
consistently produced low cross-validation error; models in the
lower-left corner had neither consistently low cross-validation error
nor consistently high cross-validation error; models in the upper-left
corner had consistently high cross-validation
error.}\label{fig:cvgoodbad}
\end{figure}
}

Finally, Appendix Figure~\ref{fig:cvgoodbadcountry} breaks the 5-fold
cross validation results down by country, for the five countries that
have data for more than 10 cohorts per sex. This figure is similar in
spirit to Figure~\ref{fig:deltaaicgoodbadcountry}, but it is based on a
different metric and thus not directly comparable. The figure shows that
5-fold cross validation results differ by country, but differences using
this summary plot are less salient than they were using
\(\Delta\text{AIC}\). The most distinctive feature of France and Italy
is that they both show the Makeham and Gompertz model consistently
performing poorly, relative to the other models.

\hypertarget{fig:cvgoodbadcountry}{%
\begin{figure}
\centering
\includegraphics[width=\textwidth,height=1.1\textwidth]{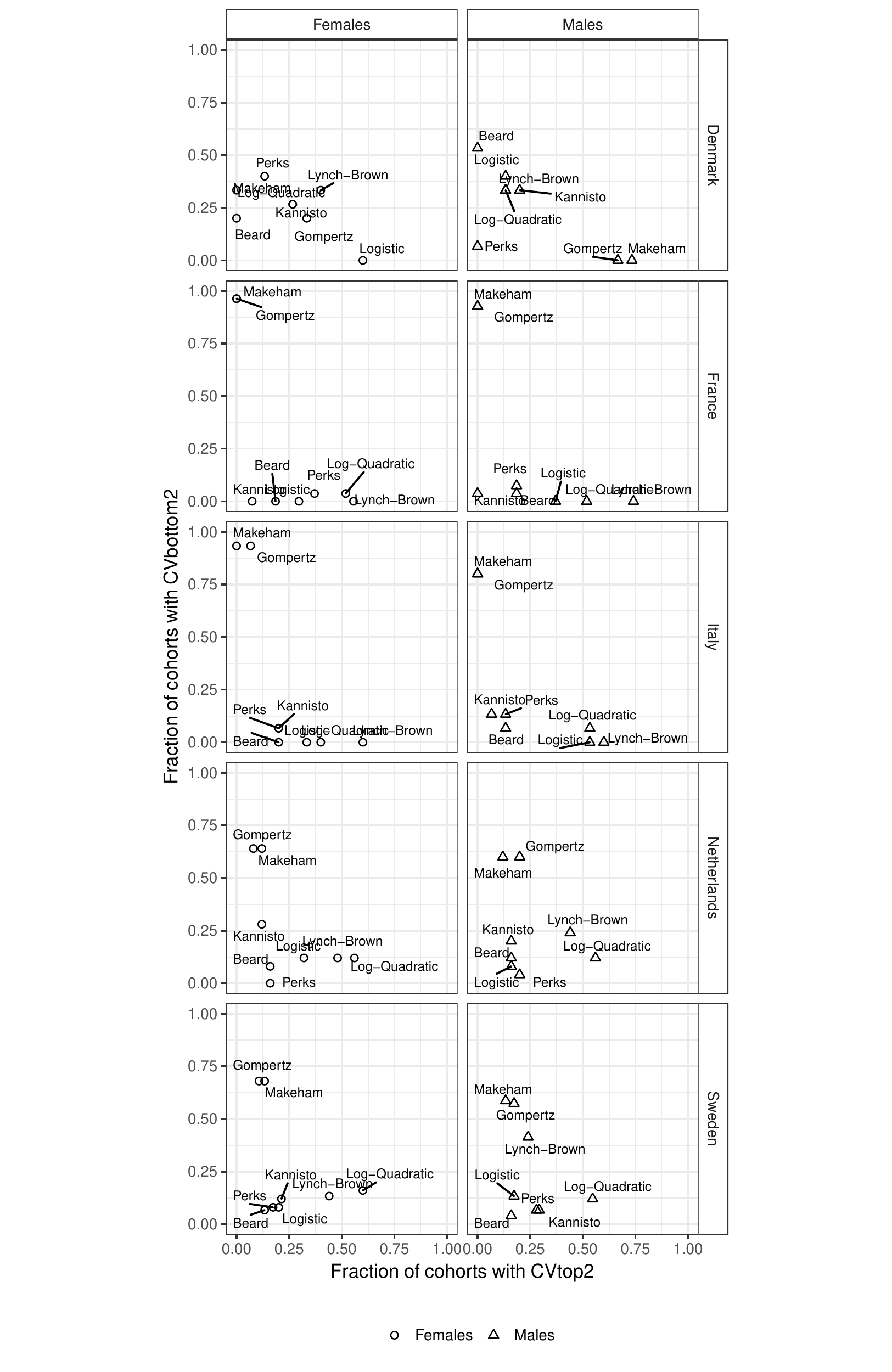}
\caption{Fraction of cohorts for which the cross-validation error of
each model was very good (best or second-best, x axis) and very poor
(worst or second-worst, y axis) by country. Models in the lower-right
corner consistently produced low cross-validation error; models in the
lower-left corner had neither consistently low cross-validation error
nor consistently high cross-validation error; models in the upper-left
corner had consistently high cross-validation
error.}\label{fig:cvgoodbadcountry}
\end{figure}
}

\newpage\clearpage

\hypertarget{sec:samplesize}{%
\section{Illustrative simulation: sample size}\label{sec:samplesize}}

We found that patterns of model fit varied by country, and that this
variation was related to how poorly the Gompertz/Makeham models fit
cohorts from each country. At one extreme, the Gompertz/Makeham models
did not fit France and Italy well at all; at the other extreme, the
Gompertz/Makeham models did not produce poor predictions for Denmark
(though they were not the best performers). One possible explanation for
this difference is sample size: cohorts from Italy and France are
roughly 5 to 10 times larger than cohorts from Sweden, Denmark, and the
Netherlands (Table~\ref{tbl:kt-data}). Thus, in a small country like
Denmark, it may be the case that small sample sizes do not provide
enough information to reliably detect non-Gompertz patterns in old-age
death rates, even if the underlying mortality process is
non-Gompertzian.

Although we cannot examine how model fits would be affected if we
observed bigger cohorts from Denmark, we can examine how our inferences
about bigger countries would be affected if we observed smaller cohorts.
To investigate this question, we started with the cohort of French
Females born in 1872 and successively reduced the cohort size. At each
reduced size, we re-fit all of the models and calculated
\(\Delta\text{AIC}\). Figure~\ref{fig:samplesize} shows that, as
expected, reducing the size of the cohorts led simpler, two parameter
models to perform relatively better. However, the Gompertz model never
emerges as the best; instead, in the regime where two-parameter models
are favored (below about 30\% of the original cohort size), the Kannisto
model -- which has a sigmoid shape, allowing for decelerating death
rates -- is the one that emerges as the best. This suggests that, at
least for France, if cohorts had been smaller but generated by the same
underlying mortality dynamics, we may have found that the Kannisto model
was most accurate, but we would not have selected the Gompertz model.

\hypertarget{fig:samplesize}{%
\begin{figure}
\centering
\includegraphics{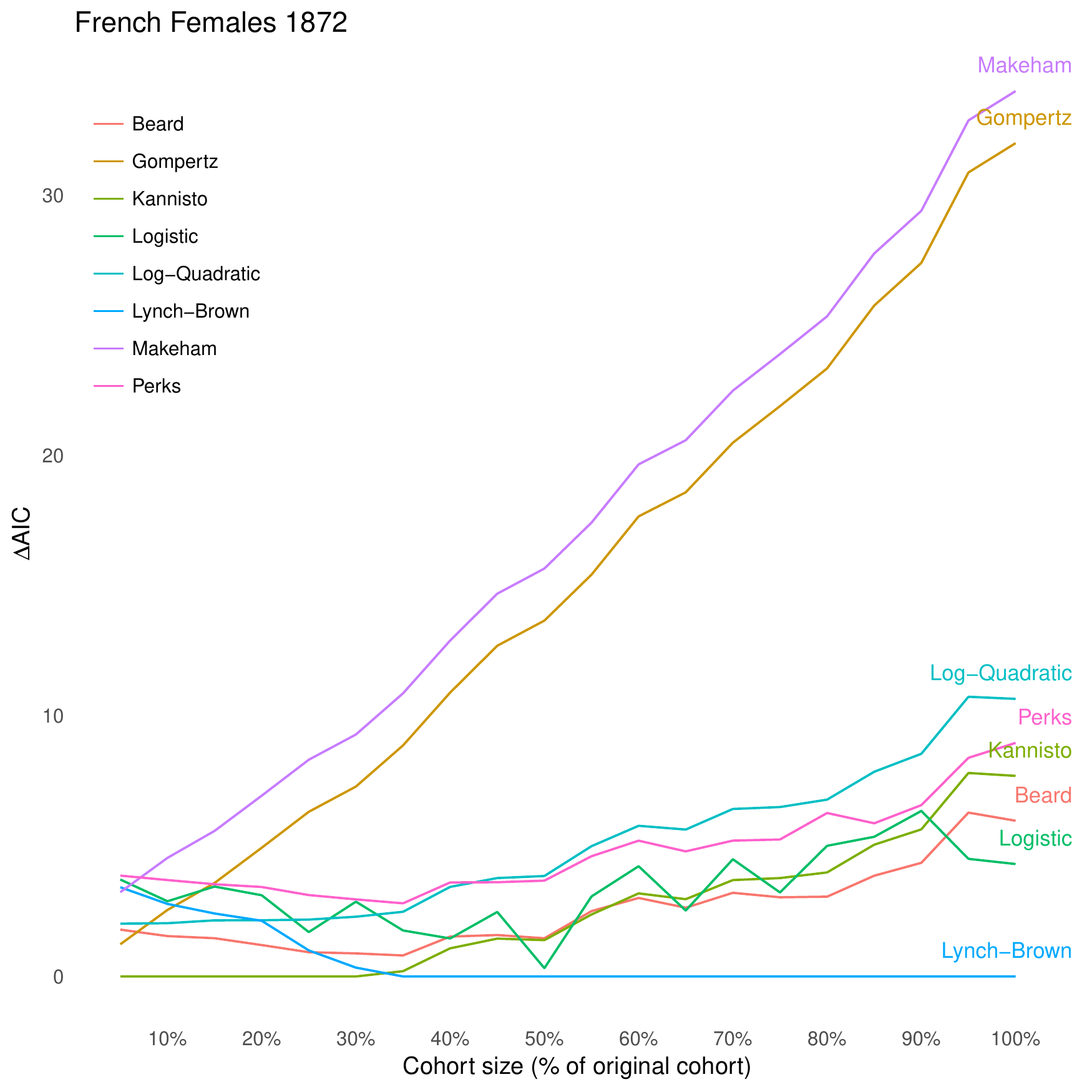}
\caption{Simulation showing the effect of reducing cohort size on
\(\Delta\text{AIC}\) using data from French Females in 1872. The
Lynch-Brown model fits the original cohort best. However, as the size of
the cohort is reduced, simpler models perform better; at around 30\% of
the original sample size, the two-parameter Kannisto model is estimated
to fit better than the four-parameter Lynch-Brown model. The Gompertz
model improves with reduced sample size, but it never surpasses the
sigmoid-shaped Kannisto model.}\label{fig:samplesize}
\end{figure}
}

\newpage\clearpage

\hypertarget{sec:hazards}{%
\section{Models}\label{sec:hazards}}

This appendix contains detailed information about each model, including
some references to the literature and details about how the model was
parameterized and fit.

\hypertarget{gompertz}{%
\subsection{Gompertz}\label{gompertz}}

The Gompertz function is one of the oldest and most common parametric
description of death rates in demography. It was first introduced in
Gompertz (1825) and its simplicity and interpretability have meant that
it continues to be used today (S. Preston et al. 2000).

\subsubsection*{Hazard function}

The Gompertz hazard can be written

\[
  \mu(z) = \alpha \exp (\beta z),
\]

where \(\alpha > 0\).

\hypertarget{parameters}{%
\subsubsection{Parameters}\label{parameters}}

The paramaterization used in the \texttt{mortfit} package is given in
Table \ref{tab:gompertzparamtab}.

\begin{table}[h]
\begin{center}
\begin{tabular}{ l c }
model & code\\
\hline
$\alpha \in (0, \infty)$ & $\exp(\theta_1)$\\
$\beta \in (-\infty, \infty)$& $\theta_2$
\end{tabular}
\caption{The parameterization of the Gompertz hazard used in the \texttt{mortfit} package.}
\label{tab:gompertzparamtab}
\end{center}
\end{table}

\subsubsection*{Starting values}

In order to choose the starting values for the BFGS optimization
algorithm, we estimate a regression of the log central death rates on
age; that is, we fit

\begin{equation}
\log(M_z) = \alpha_0 + \beta_0 z + \epsilon.
\end{equation}

We then use the preliminary estimates of \(\alpha_0\) and \(\beta_0\) as
our starting values.

\subsubsection*{Conditional probability of death}

\begin{equation}
{}_1q_z = 
1 - \exp\left(-\int_{z}^{z+1} \mu(x)dx\right) = 
1 - \exp\left( -\frac{\alpha  }{\beta }\left(e^{\beta  (z+1)}-e^{\beta z}\right) \right)
\end{equation}

\newpage\subsection{Makeham}
\label{ap:makeham}

Makeham (1860) proposed adding an additional parameter to the Gompertz
function; this additional parameter is often interpreted as a baseline
level of extrinsic mortality.

\subsubsection*{Hazard function}

The Makeham hazard function can be written

\begin{equation}
\mu(z) = \gamma + \alpha \exp(\beta z).
\end{equation}

\subsubsection*{Parameters}

\begin{table}[h]
\begin{center}
\begin{tabular}{ l c }
model & code\\
\hline
    $\alpha \in (0,\infty)$ & $\exp(\theta_1)$\\
$\beta \in (-\infty,\infty)$ & $\theta_2$\\
    $\gamma \in (0, \infty)$ & $\exp(\theta_3)$
\end{tabular}
\caption{The parameterization of the Makeham hazard used in the \texttt{mortfit} package.}
\label{tab:makehamparamtab}
\end{center}
\end{table}

\subsubsection*{Starting values}

In order to choose starting values for the BFGS algorithm, we follow a
procedure similar to the one we used for the Gompertz model. First, we
choose \(\gamma_0 = \exp(\zeta)\), where \(\zeta\) is a very small
value. Next, we estimate a regression of the log central death rates on
age; that is, we fit

\begin{equation}
\log(M_z - \gamma_0) = b_0 + b_1 z + \epsilon.
\end{equation}

For the starting parameter estimates we use \(\alpha_0 = \exp(b_0)\) and
\(\beta_0 = b_1\).

\subsubsection*{Conditional probability of death}

\begin{equation}
{}_1q_z = 
1 - \exp\left(-\int_{z}^{z+1} \mu(x)dx\right) = 
1 - \exp \left(
-\frac{\alpha  \left(e^{\beta  (z+1)}-e^{\beta  z}\right)}{\beta } - \gamma
\right)
\end{equation}

\newpage\subsection{Logistic}
\label{ap:logistic}

We follow A. Thatcher et al. (1998) in calling this the logistic hazard,
though several other hazard functions have logistic shapes. The study of
logistic-shaped hazard functions, including (but not limited to) this
one has a long history; starting with Perks (1932), and including many
subsequent studies (R. E Beard 1959,L. A Gavrilov and Gavrilova (1991),
A. Yashin et al. (1994), Le Bras (1976), Vaupel et al. (1979)).

Note that when \(\delta = 0\), this particular logistic form simplifies
to be Makeham.

\subsubsection*{Hazard function}

The Perks-Beard logistic hazard is given by

\begin{equation}
\mu(z) = \gamma + \frac{\alpha \exp(\beta z)}{1 + \delta \exp(\beta z)},
\end{equation}

where \(\alpha > 0\), \(\beta > 0\), \(\gamma > 0\), and \(\delta > 0\).

\subsubsection*{Parameters}

\begin{table}[h]
\begin{center}
\begin{tabular}{ l c }
model & code\\
\hline
$\alpha \in (0,\infty)$ & $\exp(\theta_1)$\\
$\beta \in (0,\infty)$ & $\exp(\theta_2)$\\
$\gamma \in (0, \infty)$ & $\exp(\theta_3)$\\
$\delta \in (0, \infty)$ & $\exp(\theta_4)$
\end{tabular}
\caption{The parameterization of the Logistic hazard used in the \texttt{mortfit} package.}
\label{tab:logisticparamtab}
\end{center}
\end{table}

A. Thatcher et al. (1998) reparameterized this using

\begin{align}
\alpha &= \frac{ba}{b - \sigma^2 a}\\
\beta &= b\\
\gamma &= c\\
\delta &= \frac{b}{\sigma^2a} - 1
\end{align}

\noindent to yield the expression

\begin{equation}
\mu(z | a, b, c, d) = c + \frac{a \exp(bz)}{1 + \sigma^2 \frac{a}{b}\left(\exp(bz)-1\right)}.
\end{equation}

\noindent This is useful in studies of heterogeneity; however, in this
analysis we employ the original parameterization.

\subsubsection*{Starting values}

Choosing starting values for the BFGS optimization algorithm is somewhat
tricky. First, we summarize a few derivations that are used in our
approach. Analytically, the inflection point in the logsitic curve---the
point at which the second derivative with respect to age is equal to
0---will be attained when

\begin{equation}
  z_{\text{inf}} = -\frac{\log(\delta)}{\beta}.
\label{eq:logistic-infpt}\end{equation}

\noindent Furthermore, the slope at the inflection point will be

\[
\frac{d \mu(z)}{d z} \bigg|_{z=z_{\text{inf}}} = \mu^{\prime}(z_{\text{inf}}) 
                  = \frac{\alpha \beta}{4 \delta}.
\]

\noindent This means that we have

\begin{equation}
    \beta = \frac{4 \delta \mu^{\prime}(z_{\text{inf}})}{\alpha}.  
\label{eq:logistic-betaexpr}\end{equation}

In order to choose starting values for the optimization routine, we
start by using the observed central death rates \(M_z\) as preliminary
estimates. We fit a logistic regression of the observed central death
rates on age, with each age weighted by the amount of exposure we
observe. Following a procedure similar to the one described for the
Lynch-Brown form, we use the fitted values from the model to estimate
the age at which the median of the range of observed central death rates
is attained; this is \(z_{\text{inf}}\), our crude estimate of the
inflection point in the central death rates.

Now note that \begin{equation}
\lim_{z \rightarrow \infty} \mu(z) = \frac{\alpha}{\delta} + \gamma.
\label{eq:logistic-upperlim}\end{equation}

\noindent Since we expect \(\gamma\) to be small,
\(\lim_{z \rightarrow \infty} \mu(z) \approx \frac{\alpha}{\delta}\).
Therefore, for our starting point \(\beta_0\), we combine
Equation~\ref{eq:logistic-betaexpr} and
Equation~\ref{eq:logistic-upperlim} to make the crude estimate
\(\beta_0 = 4~\mu^{\prime}(z_{\text{inf}}) / M_\text{max}\), where
\(M_\text{max}\) is the maximum of the observed central death rates.
Then, using Equation~\ref{eq:logistic-infpt}, we set the starting point
for \(\delta\) to be \(\delta_0 = \exp(-\beta/z_{\text{inf}})\).

Next, again taking advantage of the fact that \(\gamma\) is typically
very small, and using Equation~\ref{eq:logistic-upperlim}, we set the
starting point for \(\alpha\) to be
\(\alpha_0 = \delta_0 M_{\text{max}}\).

Finally, note that \begin{align}
  \lim_{z \rightarrow 0} \mu(z) = \gamma + \frac{\alpha}{1 + \delta}.
\end{align}

We use this limit to choose a starting point for \(\gamma\):
\(\gamma_0 = M_{\text{min}} - \frac{\alpha_0}{1 + \delta_0}\), where
\(M_{\text{min}}\) is the smallest observed central death rate in the
youngest age groups.

Like the starting points we use for all of the functional forms, the
main goal here is to be sure the optimization routine starts with very
approximately reasonable parameter values.

\subsubsection*{Conditional probability of death}

\begin{align}
{}_1q_z &= 
1 - \exp\left(-\int_{z}^{z+1} \mu(x)dx\right)\\ 
&= 
1 - \exp\left(
-\frac{\alpha  \left[\log \left(\beta \delta e^{\beta (z+1)}+\beta\right)-
                     \log \left(\beta \delta e^{\beta z}+\beta\right)
              \right]
      }
      {\beta  \delta } 
   - \gamma
\right)\\
&= 
1 - \exp\left(-\gamma\right)
\exp\left(
-\frac{\alpha  \left[\log \left(\beta \delta e^{\beta (z+1)}+\beta\right)-
                     \log \left(\beta \delta e^{\beta z}+\beta\right)
              \right]
      }
      {\beta  \delta } 
\right)\\
&= 
1 - \exp\left(-\gamma\right)
\exp\left(
\log \left(\beta \delta e^{\beta (z+1)}+\beta\right)-
\log \left(\beta \delta e^{\beta z}+\beta\right)
              \right)^{-\frac{\alpha}{\beta \delta}}\\
&= 
1 - \exp\left(-\gamma\right)
\left(
\frac{\beta \delta e^{\beta (z+1)}+\beta}
     {\beta \delta e^{\beta z}+\beta}
\right)^{-\frac{\alpha}{\beta \delta}}\\
&= 
1 - \exp\left(-\gamma\right)
\left(
\frac{\delta e^{\beta (z+1)} + 1}
     {\delta e^{\beta z} + 1}
\right)^{-\frac{\alpha}{\beta \delta}}\\
&= 
1 - \exp\left(-\gamma\right)
\left(
\frac{\delta e^{\beta z} + 1}
     {\delta e^{\beta (z+1)} + 1}
\right)^{\frac{\alpha}{\beta \delta}}.
\end{align}

\newpage\subsection{Kannisto}
\label{ap:kannisto}

The Kannisto model is a special case of the Logistic hazard where
\(\alpha = \delta\). A. Thatcher et al. (1998) trace its origins to a
presentation given by Kannisto in 1992, and also to Himes et al. (1994).
A. Thatcher et al. (1998) concluded that the Kannisto model, along with
the Logistic model, was the best of the set they tested.

\subsubsection*{Hazard function}

The Kannisto hazard can be written

\begin{equation}
\mu(z) = \frac{\alpha \exp(\beta z)}{1 + \alpha \exp(\beta z)},
\end{equation}

where \(\alpha > 0\) and \(\beta > 0\).

\subsubsection*{Parameters}

\begin{table}[h]
\begin{center}
\begin{tabular}{ l c }
model & code\\
\hline
$\alpha \in (0,\infty)$ & $\exp(\theta_1)$\\
$\beta \in (0,\infty)$ & $\exp(\theta_2)$
\end{tabular}
\caption{The parameterization of the Kannisto hazard used in the \texttt{mortfit} package.}
\label{tab:kannistoparamtab}
\end{center}
\end{table}

\subsubsection*{Starting values}

In order to choose the starting values for the BFGS optimization
algorithm, we estimate a logistic regression of the central death rates
on age; that is, we use \texttt{R}'s \texttt{glm} function to fit:

\begin{equation}
\text{logit}^{-1}(M_z) = a + b z.
\end{equation}

We then use the coefficient estimates from the regression for starting
values: for \(\alpha\), we use \(\alpha_0 = \exp(a)\) (equivalently,
because of our parameterization, \(\theta_1\) starts at \(a\)); and for
\(\beta\), we use \(\beta_0 = b\) (equivalently, because of our
parameterization, \(\theta_2\) starts at \(\log b\)).

\subsubsection*{Conditional probability of death}

\begin{align}
{}_1q_z &= 
1 - \exp\left(-\int_{z}^{z+1} \mu(x)dx\right)\\
&= 1 - \exp\left(-
\frac{\log \left(\alpha \beta e^{\beta  (z+1)} + \beta\right)-
      \log \left(\alpha \beta e^{\beta z} + \beta\right)}{\beta }
   \right)\\
&= 1 - \exp\left( \left[
      \log \left(\alpha \beta e^{\beta z} + \beta\right)-
  \log \left(\alpha \beta e^{\beta  (z+1)} + \beta\right) \right]^{1/\beta}
      \right)\\
&= 1 - \exp\left(\log\left(
          \frac{\alpha \beta e^{\beta z} + \beta}
          {\alpha \beta e^{\beta(z+1)} + \beta} 
            \right)^{1/\beta} \right)\\
&= 1 - \left(
          \frac{\alpha  e^{\beta z} + 1}
          {\alpha e^{\beta(z+1)} + 1} 
            \right)^{1/\beta}\\
\end{align}

\newpage 

\hypertarget{beard-gamma-gompertz}{%
\subsection{Beard / Gamma-Gompertz}\label{beard-gamma-gompertz}}

The Beard model is a type of logistic hazard that can be derived from a
model in which (1) individuals share a common baseline hazard and (2)
each individual has a fixed frailty parameter which is multiplied by the
baseline hazard to produce the individual hazard; and (3) these
individual fixed frailty parameters have a Gamma distribution. This
model has been widely studied; see, for example, R. E Beard (1959), K.
G. Manton et al. (1981), and Horiuchi and Wilmoth (1998).

\hypertarget{hazard-function}{%
\subsubsection{Hazard function}\label{hazard-function}}

The Beard hazard has the form

\begin{equation}
\mu(z) = \frac{\alpha \exp(\beta z)}{1 + \delta \exp(\beta z)},
\end{equation}

where \(\alpha > 0\), \(\beta > 0\), and \(\delta > 0\).

\hypertarget{parameters-1}{%
\subsubsection{Parameters}\label{parameters-1}}

\begin{table}[h]
\begin{center}
\begin{tabular}{ l c }
model & code\\
\hline
$\alpha \in (0,\infty)$ & $\exp(\theta_1)$\\
$\beta \in (0,\infty)$ & $\exp(\theta_2)$\\
$\delta \in (0,\infty)$ & $\exp(\theta_3)$
\end{tabular}
\caption{The parameterization of the Beard hazard used in the \texttt{mortfit} package.}
\label{tab:beardparamtab}
\end{center}
\end{table}

\hypertarget{starting-values}{%
\subsubsection{Starting values}\label{starting-values}}

We adopt the same approach used for the logistic function, except that
we do not need starting values for \(\gamma\), since that parameter is
not in the Beard hazard.

\hypertarget{conditional-probability-of-death}{%
\subsubsection{Conditional probability of
death}\label{conditional-probability-of-death}}

The conditional probability of death for the Beard hazard is the same as
for the logistic hazard with \(\gamma\) fixed at 0. Thus,

\begin{align}
{}_1q_z &= 
1 - \exp\left(-\int_{z}^{z+1} \mu(x)dx\right)\\
&= 
1 - 
\left(
\frac{\delta e^{\beta z} + 1}
     {\delta e^{\beta (z+1)} + 1}
\right)^{\frac{\alpha}{\beta \delta}}.
\end{align}

\newpage

\hypertarget{perks-gamma-makeham}{%
\subsection{Perks / Gamma-Makeham}\label{perks-gamma-makeham}}

The Perks model is another type of logistic hazard. While the Beard
population hazard can be derived from a model that assumes a
Gamma-Gompertz distribution of individual hazards, the Perks model can
be derived from a model that assumes a Gamma-Makeham distribution of
individual hazards. See, for example, K. G. Manton et al. (1981), and
Horiuchi and Wilmoth (1998).

\hypertarget{hazard-function-1}{%
\subsubsection{Hazard function}\label{hazard-function-1}}

The Perks hazard has the form

\begin{equation}
\mu(z) = \frac{\gamma + \alpha \exp(\beta z)}{1 + \delta \exp(\beta z)},
\end{equation}

where \(\alpha > 0\), \(\beta > 0\), \(\gamma > 0\), and \(\delta > 0\).

\hypertarget{parameters-2}{%
\subsubsection{Parameters}\label{parameters-2}}

\begin{table}[h]
\begin{center}
\begin{tabular}{ l c }
model & code\\
\hline
$\alpha \in (0,\infty)$ & $\exp(\theta_1)$\\
$\beta \in (0,\infty)$ & $\exp(\theta_2)$\\
$\gamma \in (0,\infty)$ & $\exp(\theta_3)$\\
$\delta \in (0,\infty)$ & $\exp(\theta_4)$
\end{tabular}
\caption{The parameterization of the Perks hazard used in the \texttt{mortfit} package.}
\label{tab:beardparamtab}
\end{center}
\end{table}

\hypertarget{starting-values-1}{%
\subsubsection{Starting values}\label{starting-values-1}}

We use the same approach described for the Logistic function to choose
starting values when fitting the Perks hazard.

\hypertarget{conditional-probability-of-death-1}{%
\subsubsection{Conditional probability of
death}\label{conditional-probability-of-death-1}}

The conditional probability of death for the Perks hazard is

\begin{align}
{}_1q_z &= 
1 - \exp\left(-\int_{z}^{z+1} \mu(x)dx\right)\\
&= 
1 -
\exp\left(-\gamma + \frac{\gamma \log\left(\delta e^{\left(\beta x + \beta\right)} + 1\right)}{\beta} - \frac{\gamma \log\left(\delta e^{\left(\beta x\right)} + 1\right)}{\beta} - \frac{\alpha \log\left(\delta e^{\left(\beta x + \beta\right)} + 1\right)}{\beta \delta} + \frac{\alpha \log\left(\delta e^{\left(\beta x\right)} + 1\right)}{\beta \delta}\right)\\
&= 1 - \exp(-\gamma)
\left[ \frac{\delta \exp(\beta x + \beta) + 1}{\delta \exp(\beta x) + 1} \right]^{\frac{\gamma}{\beta}}
\left[ \frac{\delta \exp(\beta x) + 1}{\delta \exp(\beta x + \beta) + 1} \right]^{\frac{\alpha}{\beta \delta}}.
\end{align}

\newpage\subsection{Weibull}
\label{ap:weibull}

The Weibull hazard is commonly used in engineering to describe the
failure of systems composed of many components; L. A Gavrilov and
Gavrilova (1991) has a good discussion of the Weibull model, though
those authors did not recommend it as a model of human mortality.

\subsubsection*{Hazard function}

The Weibull hazard has the form

\begin{equation}
\mu(z) = \alpha z^{\beta-1}.
\end{equation}

See, for example, K. G. Manton et al. (1986).

\subsubsection*{Parameters}

\begin{table}[h]
\begin{center}
\begin{tabular}{ c c }
model & code\\
\hline
$\alpha \in (0,\infty)$ & $\exp(\theta_1)$\\
$\beta \in (0,\infty)$ & $\exp(\theta_2)$
\end{tabular}
\caption{The parameterization of the Weibull hazard used in the \texttt{mortfit} package.}
\label{tab:weibullparamtab}
\end{center}
\end{table}

\subsubsection*{Starting values}

In order to compute starting values for the BFGS optimization method, we
fit a regression of the log of the central death rates on the log of
age; that is, we fit

\begin{equation}
\log(M_z) = a + b \log(z).
\end{equation}

We then use \(\alpha_0 = a\) as the initial estimate of \(\alpha\), and
\(\beta_0 = \log(b)-1\) as the initial estimate of \(\beta\).

\subsubsection*{Conditional probability of death}

\begin{equation}                                                                            
{}_1q_z =                                                                                   
1 - \exp \left(-\int_{z}^{z+1} \mu(x)dx\right) =                                            
1 - \exp \left(\frac{\alpha}{\beta} \left( z^{\beta} - (z+1)^{\beta} \right) \right).
\end{equation}

\newpage\subsection{Log-Quadratic}
\label{sec:logquadratic}

This hazard function is interesting in part because it can produce
hazards that decelerate at oldest ages, but it is not logistic. It has
been studied by several authors, including Horiuchi (2003) and
Steinsaltz and Wachter (2006).

\subsubsection*{Hazard function}

The log-quadratic hazard function has the form

\begin{equation}
\mu(z) = \exp\left(\alpha + \beta z + \gamma z^2\right).
\end{equation}

\subsubsection*{Parameters}

\begin{table}[h]
\begin{center}
\begin{tabular}{ l c }
model & code\\
\hline
$\alpha \in (-\infty,\infty)$ & $\theta_1$\\
$\beta \in (-\infty,\infty)$ & $\theta_2$\\
$\gamma \in (-\infty, \infty)$ & $\theta_3$
\end{tabular}
\caption{The parameterization of the Log-quadratic hazard used in the \texttt{mortfit} package.}
\label{tab:logquadraticparamtab}
\end{center}
\end{table}

\subsubsection*{Starting values}

We fit a regression of the logged crude death rates on age and age
squared. The estimated coefficients from that regression are the
starting values for the BFGS algorithm.

\subsubsection*{Conditional probability of death}

\begin{align}
{}_1q_z 
&= 1 - \exp\left(-\int_{z}^{z+1} \mu(x)dx\right)\\
&= 
1
-\exp\left(\frac{\sqrt{\pi} \text{erf}\left(\frac{2 \, \gamma x + \beta + 2 \, \gamma}{2 \, \sqrt{-\gamma}}\right) e^{\left(\alpha - \frac{\beta^{2}}{4 \, \gamma}\right)}}{2 \, \sqrt{-\gamma}} - \frac{\sqrt{\pi} \text{erf}\left(\frac{2 \, \gamma x + \beta}{2 \, \sqrt{-\gamma}}\right) e^{\left(\alpha - \frac{\beta^{2}}{4 \, \gamma}\right)}}{2 \, \sqrt{-\gamma}}\right)\\
&= 
1
-\exp\left(
\frac{\sqrt{\pi} e^{\left(\alpha - \frac{\beta^{2}}{4 \, \gamma}\right)}}{2 \, \sqrt{-\gamma}} 
\left[
\text{erf}\left(\frac{2 \, \gamma x + \beta + 2 \, \gamma}{2 \, \sqrt{-\gamma}}\right)
- 
\text{erf}\left(\frac{2 \, \gamma x + \beta}{2 \, \sqrt{-\gamma}}\right)
\right]
\right)\\
&= 
1
-\exp\left(
\frac{\sqrt{\pi} e^{\left(\alpha - \frac{\beta^{2}}{4 \, \gamma}\right)}}{2 \, \sqrt{-\gamma}} 
\left[
\text{erf}\left(\frac{\beta + 2 \, \gamma (x + 1)}{2 \, \sqrt{-\gamma}}\right)
- 
\text{erf}\left(\frac{\beta + 2 \, \gamma x}{2 \, \sqrt{-\gamma}}\right)
\right]
\right)
\end{align}

\noindent where erf is the error function.

\newpage\subsection{Lynch-Brown}
\label{ap:lb}

\subsubsection*{Hazard function}

The Lynch-Brown hazard function was developed to naturally parameterize
deceleration and compression in mortality hazards (S. M. Lynch and Brown
2001). The Lynch-Brown hazard is given by

\begin{equation}
\mu(z) = \alpha + \beta \text{arctan} \{\gamma(z - \delta) \}.
\end{equation}

For given values of \(\beta, \gamma\) the hazard yields the same hazard
as with \(-\beta,-\gamma\). In order to identify it, S. M. Lynch and
Brown (2001) proposes the constraint that \(\beta > 0\). We also use
this constraint.

\subsubsection*{Parameters}

\begin{table}[h]
\begin{center}
\begin{tabular}{ l c }
model & code\\
\hline
$\alpha \in (-\infty,\infty)$ & $\theta_1$\\
$\beta \in (0,\infty)$ & $\exp(\theta_2)$\\
$\gamma \in (0, \infty)$ & $\exp(\theta_3)$\\
$\delta \in (-\infty, \infty)$ & $\theta_4$
\end{tabular}
\caption{The parameterization of the Lynch-Brown hazard used in the \texttt{mortfit} package.}
\label{tab:lbparamtab}
\end{center}
\end{table}

S. M. Lynch and Brown (2001) has a useful explication of the
interpretation of each of these parameters. Overall the hazard has a
sigmoid shape. \(\alpha\) and \(\delta\) are shape parameters.
\(\alpha\) shifts the inflection point of the sigmoid shape up and down.

\(\beta\) is a scale parameter, controlling the height of the sigmoid
shape. Larger values of \(\beta\) imply a greater range of hazard values
across ages.

\(\gamma\) stretches the curve horizontally, with larger values of
\(\gamma\) making the hazard traverse the middle range of the sigmoid
shape more quickly.

\(\delta\) indicates the age at which the inflection point of the
sigmoid shape is attained. We allow it to be negative, since we
parameterize age 80 as \(z=1\) and it's possible that the inflection
point could be estimated to be attained before age 80.

\subsubsection*{Fit details}

Table 1 of S. M. Lynch and Brown (2001) shows parameter estimates from
the Lynch-Brown hazard fit to data in the US. We choose the baseline
year estimates and take the range of reasonable parameter values to be
these plus and minus 3 standard deviations. Since we paramaterize age
\(z=1\) to be 80 years, we subtract 79 from the reported \(\delta\)
estimates.

Choosing starting values for the BFGS optimization algorithm is not
trivial. We have found that the following procedure has yielded good
results: first, compute central death rates
\(M_z = \frac{D_z}{N_z - 0.5 D_z}\). As S. M. Lynch and Brown (2001)
point out, the \(\beta\) parameter is the maximum value that the hazard
attains, scaled by \(\pi\), so we take the initial value of \(\beta\) to
be \(\frac{\max_{z} D_z}{\pi}\).\\
The \(\delta\) parameter is the age at which the inflection point
occurs; for an initial estimate, we determine the age at which the
fitted values from the logistic regression are closest to the midpoint
of the observed central death rates, and take \(\delta\) to be that
value.\\
\(\gamma\) is the slope of the hazard at the inflection point; as a
starting value, we use a numerical approximation to the derivative of
the fitted logistic curve at age \(\delta\).\\
Finally, we choose \(\alpha\) so that we are guaranteed that the hazard
is positive over the range of ages we have data for.

\subsubsection*{Conditional probability of death}

\begin{align}
{}_1q_z =&
1 - \exp\left(-\int_{z}^{z+1} \mu(x)dx\right)\\
&= 1 -
\exp\left(
-\alpha -
\frac{-\beta}{2\gamma}
\left[
2 k_1 \arctan(k_1) -
2 k_0 \arctan(k_0) +
\log(k_0^2 + 1) -
\log(k_1^2 + 1)
\right]
\right),
\end{align}

where we have defined \(k_0 = \gamma(z - \delta)\) and
\(k_1 = \gamma(z - \delta + 1)\).

\newpage

\end{document}